\documentclass[10pt]{bmc_article}

\usepackage{cite} 
\usepackage{url}  
\usepackage{ifthen}  
\usepackage{multicol}   
\usepackage[utf8]{inputenc} 
\usepackage{graphicx}
\usepackage{subfigure}
\usepackage{bm,comment}
\newcommand{\remove}[1]{}
\newboolean{publ}

\newenvironment{bmcformat}{\baselineskip20pt\sloppy\setboolean{publ}{false}}{\baselineskip20pt\sloppy}

\begin{document}
\begin{bmcformat}

\title{Bounded Rationality in Scholarly Knowledge Discovery}

\author{Kristina Lerman$^1$\correspondingauthor
         \email{Kristina Lerman\correspondingauthor - lerman@isi.edu}
       and
         Nathan Hodas$^2$
         \email{Nathan Hodas -nathanhodas@pnnl.gov}
       and
       Hao Wu$^1$
         \email{Hao Wu -hwu@usc.edu}
      }
\address{%
    \iid(1) USC Information Sciences Institute,  Marina del Rey, CA, USA \\
    \iid(2) Pacific Northwest National Laboratory, Richland, WA, USA
}%

\maketitle

\begin{abstract}
In an information-rich world, people's time and attention must be divided among rapidly changing information sources and the diverse tasks demanded of them.
How people decide which of the many sources, such as scientific articles or patents, to read and use in their own work affects dissemination of scholarly knowledge and adoption of innovation.
We analyze the choices people make about what information to propagate on the citation networks of Physical Review journals, US patents and legal opinions.
We observe regularities in behavior consistent with human bounded rationality: rather than evaluate all available choices, people rely on simply cognitive heuristics to decide what information to attend to. 
We demonstrate that these heuristics bias choices, so that people preferentially propagate information that is easier to discover, often because it is newer or more popular.
However, we do not find evidence that popular sources help to amplify the spread of information beyond making it more salient.
Our paper provides novel evidence of the critical role that bounded rationality plays in the decisions to allocate attention in social communication.
\end{abstract}

\section{Introduction}
Scholarly knowledge is expressed through published works---scientific papers, books, patents, and legal opinions---which are linked together through citations. Citations analysis has emerged as a method to identify important scientific works~\cite{Redner05,chen07} and trends~\cite{Borner04,Rosvall08,Mazloumian11}, measure the productivity of individual scientists~\cite{hindex,Radicchi09}, quantify economic value of innovations~\cite{patent-value,patents}, and identify important court decisions~\cite{Fowler07,Neale13}. Citations analysis is at the core of new proposals for allocating federal research budget~\cite{Bollen13}.

Researchers, however, have questioned the objectivity of citations as a measure of importance of a scholarly work. Evidence suggests that citations may be biased by social influence~\cite{Schweitzer14}, coverage in the popular press~\cite{Phillips91}, and even the position on a web page the paper appears~\cite{arxiv-visibility}.
%
The present work extends this line of inquiry to examine the cognitive and attentional factors that play a role in the citing behavior.
Our central tenet is that knowledge discovery in citation networks is a human activity, and as such, constrained by the available time, motivation, and cognitive resources people are willing to invest in finding, processing, and evaluating information. This phenomenon, referred to as \emph{bounded rationality} by economists~\cite{Kahneman03,simon1982models}, profoundly affects the decisions people make. Because people cannot attend to all information, they employ a variety cognitive heuristics to quickly (and often unconsciously) decide what information to pay attention to~\cite{Kahneman11}. These heuristics introduce predictable biases into decision-making processes shaping human behavior, including how people discover scholarly works. Social influence, also known as the ``bandwagon effect'', is one such cognitive heuristic: people pay attention to the choices of other people.
Another important cognitive heuristic is ``position bias''~\cite{Payne51}: people pay more attention to items at the top of the list than those below them. This bias explains why information fails to spread widely in social media~\cite{Lerman2016futureinternet} and may be the reason why articles in top positions in the daily digest of papers submitted to \emph{arXiv} preprint server receive more citations~\cite{Dietrich08,arxiv-visibility}.


We investigate how bounded rationality affects citations patterns by conducting a comparative analysis of citations made by physics articles, U.S. patents and federal court decisions. The domains we study vary in the type of information citations communicate, the motivations of participants to cite, the effort they are willing to expend in knowledge discovery, and the conventions established in citing. Beyond this, they are similar in how people discover and evaluate knowledge. We describe an empirical approach that disentangles multiple factors contributing to the strategies people use to discover and link to existing information. The key to our approach is to split the data into 
more homogeneous populations 
and to carry out `human response dynamics'-based analysis within each population~\cite{Hodas12socialcom}.

Our study identifies common patterns in citing behavior. (1) We find that people pay more attention to works that are easier to find, for example, because they are cited by popular works. 
For scientific papers, this may cause some papers to receive more citations than higher quality papers that are more difficult to find.
(2) We find that attention to scholarly works decays over time in all domains: people pay more attention to more recent information~\cite{Barabasi13}.
(3) Also, other factors being equal, people pay more attention to popular works, where popularity is measured by the number of citations.
On the other hand, (4) we  found no evidence for another type of social influence: being cited by a popular work does not attract preferentially more attention to the cited work. This is contrary to the \emph{halo effect} in psychology, in which a person's ability in one area colors the perceptions of others of his or her abilities in other areas.

In order to understand and model knowledge discovery in citations networks, we must account for bounded rationality.
While our study does not replace controlled experiments, our analysis of existing behavioral data can guide future experiments to explore the causal relationships proposed here.

\section{Methods}
\label{sec:data}

\subsection{Data}
We conduct comparative empirical analysis of citations networks from three different domains.

\paragraph{Scientific papers}
Scientists communicate discoveries and research findings by publishing papers, in which they situate their work within a field by citing existing papers.
In our study, we used a large historic data set provided by the American Physical Society (APS). The APS publishes prestigious physics journals, including Physical Review Letters, the Physical Review, and Reviews of Modern Physics. The APS dataset contains  information about more than 450,000 articles published by the APS journals between the years 1893 and 2007 and the citations made by these articles to other articles in the dataset~\cite{Redner05}. A caveat about this dataset: it only includes citations to other papers published by the APS and not to papers published in other  journals. The documented citations represent the lower bound on the citations made by and to each paper, and our conclusions could be easily verified on larger universes of citation data.

\paragraph{Patents}
A patent grants an innovator an exclusive use of the proposed idea for a limited period of time. Like scientific papers, patents link to existing state-of-the-art to demonstrate the novelty of an idea or technology.
We use an historic data set of citations made by U.S. patents awarded between the years 1963 and 1999. However, these data include citations made by patents awarded between 1975 and 1999. A caveat about this data set is that patents are reviewed by patent examiners, who sometimes insert their own citations. We do not distinguish between citations made by the innovator and those inserted by the patent examiner.

\paragraph{Legal decisions}
Law is the system of rules and conventions that regulate the actions of individuals in a society. Laws make social and economic relationships predictable and consistent, yet laws evolve over time as new laws are created to reflect evolving social norms and are tested in courts. The outcome of a court case is a legal decision or an opinion.
The legal system has developed conventions for representing and communicating knowledge, utilizing citations to other judicial opinions, combined with signals of agreement/disagreement, context and specificity.

As a result, case-law, the law as determined by the courts, is encoded in the citations network.
We have used a large digitized record of federal court opinions from the CourtListener project in our study.

\subsection{Data Characteristics}

We represent each citations network as a directed graph in which nodes are documents, e.g., physics papers, and edges are citations. We use a convention in which edges run from a citing document to a cited document. Hence, outgoing edges represent citations made by a document, i.e., its bibliography,  and incoming edges represent citations of the document by other documents.

\begin{figure}[htbp] %
\centering
   \begin{tabular}{@{}c@{}c@{}c@{}}
   \includegraphics[width=0.33\textwidth]{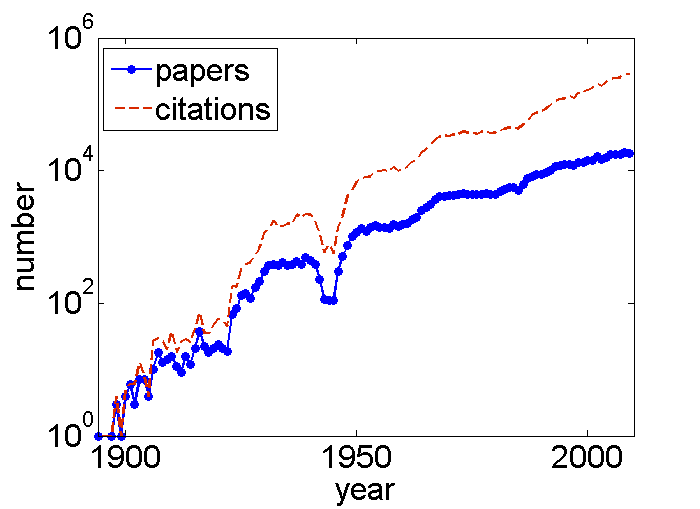} &
   \includegraphics[width=0.33\textwidth]{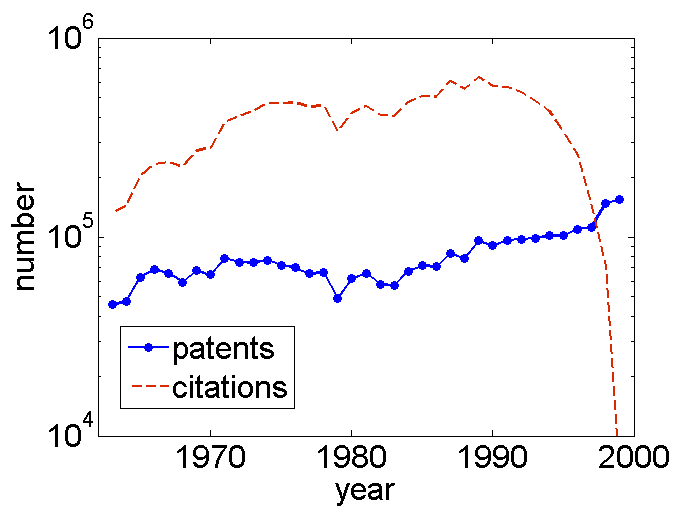} &
   \includegraphics[width=0.33\textwidth]{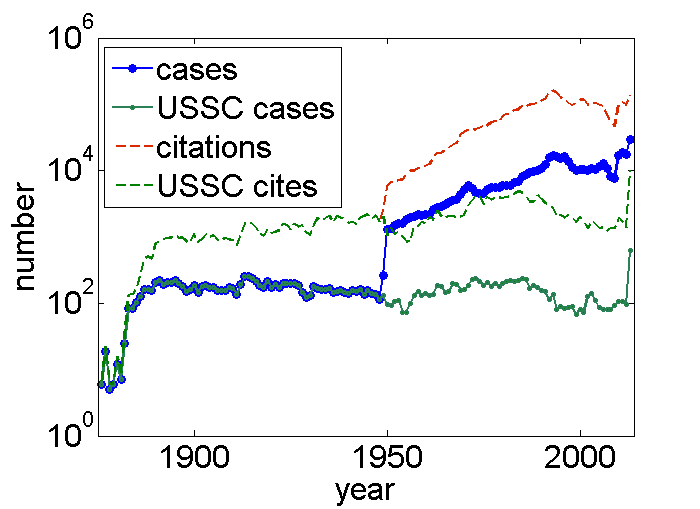} \\
(a) papers & (b) patents & (c) legal opinions
\end{tabular}
   \caption{Number of new nodes and edges in the citation network over time. }
   \label{fig:nodes}
\end{figure}

Figure~\ref{fig:nodes} shows the number of new nodes created in each citation network over time and the number of new citations made by the nodes. The dip in publication of physics papers prior to 1950 corresponds to the period of World War II. Aside from this dip, and another smaller one during World War I, the number of papers published by APS has grown significantly over the decades. The number of citations has increased even faster than the number of papers, due to an increase in the average number of references to previous works made in each paper. As in the scientific citations domain, the number of patents has also grown over the years, with the number of citations made by patents increasing even faster. In the legal citations data, we distinguish between decisions made by the Supreme Court of the United States (USSC) decisions and other federal opinions. Note that the discontinuity around 1950 corresponds to an increase in availability of digitized data and does not reflect underlying differences in the number of opinions written.

\subsection{Mutual Information}
In addition to Pearson correlation, we use mutual information to study relationships between variables. It's advantage is that it allows us to study relationships that go beyond linearity.
Mutual information $I(X;Y)$ is a non-parametric estimate of the reduction in uncertainty about a variable $X$ achieved by knowing another variable $Y$.
Large value of mutual information indicates there is a strong dependence between the variables, while $I(X;Y)=0$ means that they are independent.

We use conditional mutual information to test whether another variable $Z$ may explain the effects of $Y$ on $X$. Conditional mutual information is defined as  $I(X;Y|Z) = I(X;Y,Z) – I(X;Z)$.
If $Y$ is merely a proxy for $Z$, then $I(X;Y|Z) < I(X;Y)$, in the extreme case, when $Y$ is equal to $Z$, $I(X;Y|Z)=0$. If, on the other hand, $Y$ and $Z$ together better explain $X$ than either does alone, than $I(X;Y | Z) > I(X; Y)$.

\section{Results}



We start with comparative analysis of citation patterns across all three domains. Results reveal similarities across domains, which suggests commonalities of behaviors in citation networks.

\begin{figure}[htbp] %
\centering
   \begin{tabular}{@{}c@{}c@{}c@{}}
   \includegraphics[width=0.33\textwidth]{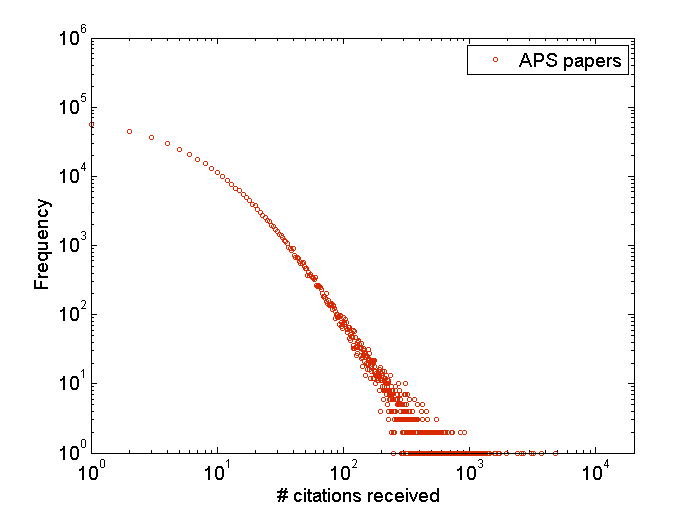} &
   \includegraphics[width=0.33\textwidth]{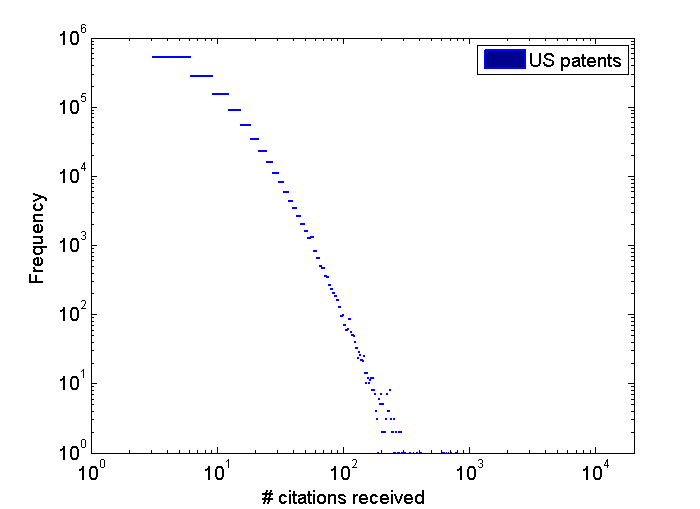} &
   \includegraphics[width=0.33\textwidth]{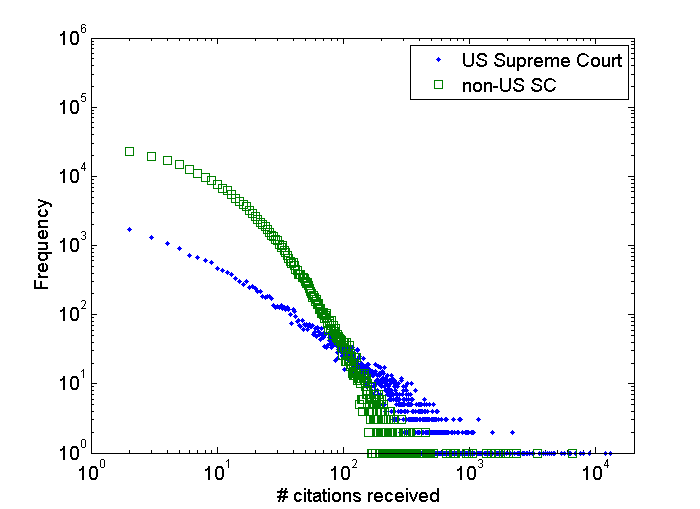} \\
(a) papers & (b) patents & (c) legal opinions
\end{tabular}
   \caption{Distribution of the number of times (a) a scientific paper, (b) a patent and (c) legal opinion were cited by other documents. }
   \label{fig:citations}
\end{figure}

\begin{figure}[htbp] %
\centering
   \begin{tabular}{@{}c@{}c@{}c@{}}
   \includegraphics[width=0.33\textwidth]{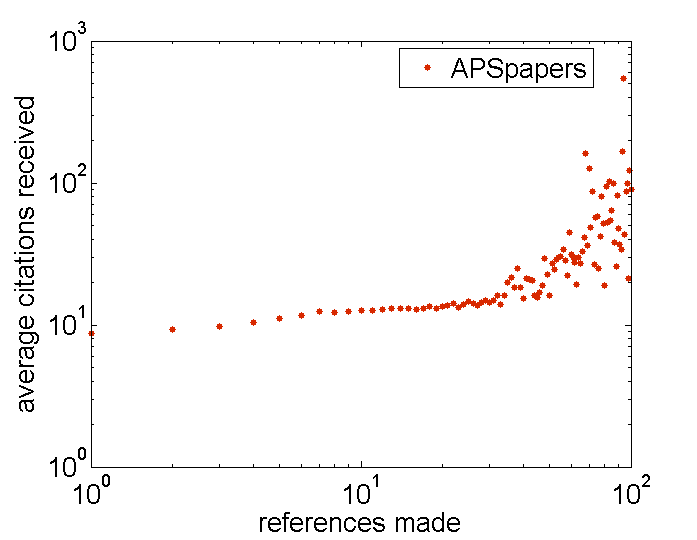} &
   \includegraphics[width=0.33\textwidth]{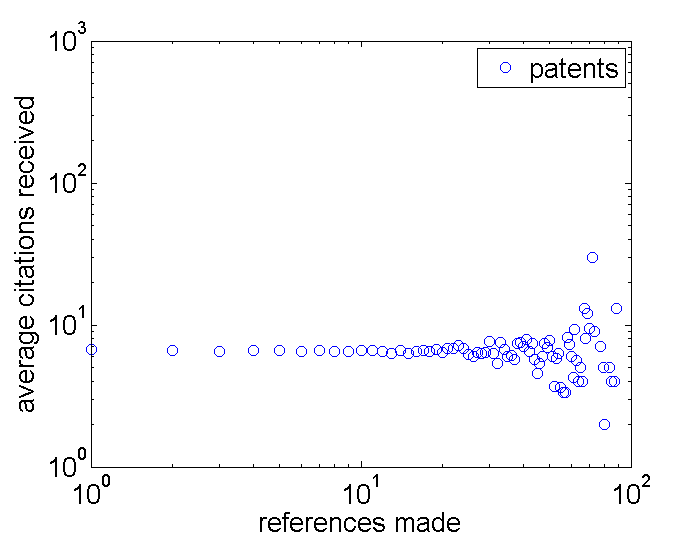} &
   \includegraphics[width=0.33\textwidth]{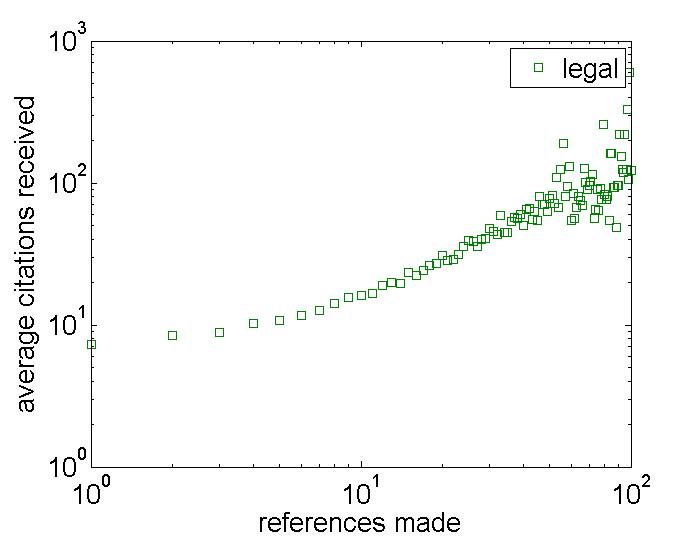} \\
(a) papers & (b) patents & (c) legal opinions
\end{tabular}
   \caption{Average number of citations as a function of the number of references made by documents. }
   \label{fig:citations_vs_refs}
\end{figure}

\subsection{Citations Inequality}
It has been known since the infancy of bibliometrics research that the recognition scientific papers receive, as measured by the number of citations, is extremely unequal, with a handful of papers receiving thousands of citations, while most of the other papers are seldom cited. In fact, almost 70\% of the papers in the APS data set were cited fewer than ten times. Figure~\ref{fig:citations}(a) shows this heterogeneous distribution.

The distribution of patent citations (Fig.~\ref{fig:citations}(b)) has a similar trend. Citations inequality is greater for patents: 85\% of the patents were cited fewer than ten times. Since citations count is thought to reflect the patents' economic value~\cite{patent-value,patents}, this result suggests that nine out of ten patents hold little value.
Figure~\ref{fig:citations}(c) shows the distribution of citations to Supreme Court opinions and those made by other courts. Compared to scientific papers and patents, legal citations have far less inequality: only about 60\% of the Supreme Court opinions receive ten or fewer citations.

What gives rise to inequality in citations? Merton~\cite{Merton68} argued that the process of ``cumulative advantage'' in science brings greater recognition to scientists who are already distinguished, resulting in unequal recognition scientists receive. Since Merton's classic work, many researchers have speculated about the origins ``cumulative advantage''~\cite{Allison82,Radicchi08unversality,Eom11}, with some attributing it to the ``first mover advantage''~\cite{Newman05} and others suggesting that limited attention is the cause~\cite{Franck99,Klamer02}.
Similar to the latter argument, we believe that inequality is a byproduct of bounded rationality. 
The cognitive heuristics that affect people's decisions to allocate their limited attention lead to inequality, which is further exacerbated by the growth of literature~\cite{Barabasi12}.

\subsection{Temporal Response}
Temporal patterns of citations hint at the processes of information discovery in these diverse domains. We examine the following temporal properties: document's age at the time of first citation, at the time of any citation, and time since its last citation.

\paragraph{Age at First Citation}

\begin{figure*}[htb] %
   \centering
   \begin{tabular}{@{}c@{}c@{}c@{}}
      \includegraphics[width=0.33\textwidth]{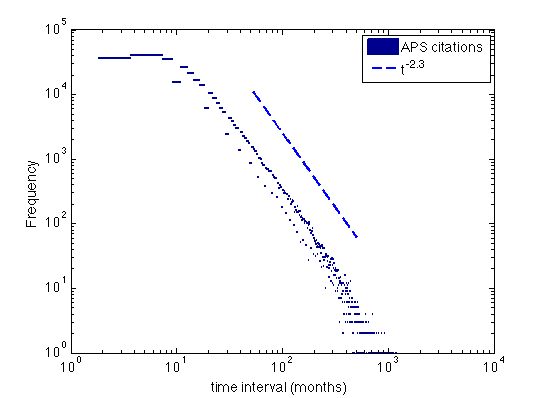} &
   \includegraphics[width=0.33\textwidth]{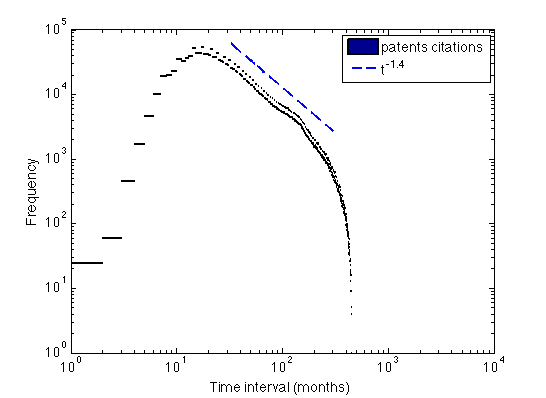} &
   \includegraphics[width=0.33\textwidth]{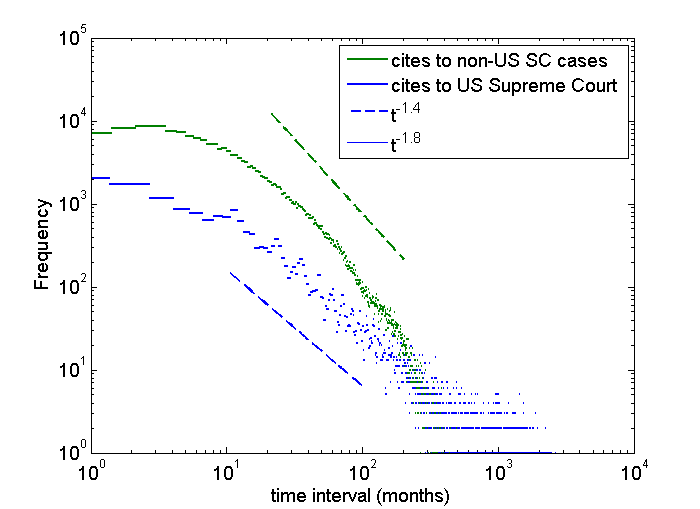}
\\
   (a) papers & (b) patents & (c) legal
   \end{tabular}
   \caption{Distribution of time to first citation of (a) scientific papers, (b) patents, and (c) U.S. Federal Court decisions. Dashed lines are guides for the eye.}
   \label{fig:first-cite-dist}
\end{figure*}

Figure~\ref{fig:first-cite-dist} shows the frequency distribution of the time to first citation of scientific papers published since 1960 (Fig.~\ref{fig:first-cite-dist}(a)) and patents granted over a similar time period (Fig.~\ref{fig:first-cite-dist}(b)). Figure~\ref{fig:first-cite-dist}(c) shows the age of a legal opinion at the time it is first cited.
These empirical results reveal the time scales of knowledge discovery in these domains. In general, the age of the document strongly decreases its chance to be discovered \emph{de novo}, even after accounting for the shrinking pool of uncited documents over time.
Unlike scientific papers, which are most likely to be cited within 8--10 months of publication, patents are most likely to be discovered within about two years of their grant date (Fig.~\ref{fig:first-cite-dist}(b)). In both data sets, the peaks correspond to the period the document is most easily discovered.  Until the rise of the web, a newly published physics paper appeared in a journal, which was easily accessible on the shelves of academic libraries. After some period of time, typically a year, journal issues were bound into volumes and moved to less accessible places in the library, requiring a greater effort for discovery. Citations peak corresponds to the period of greatest accessibility of the paper (when it is on the library shelves). The two-year peak in patent citation, on the other hand, is probably related to the time scale of patent approval. In our data set, the time interval between a patent application and approval is sharply peaked at two years. Hence, while the patent is most discoverable shortly after its publication (at the time it is granted), we do not see the evidence of the discovery until two years later, when the citing patent is granted.
In the legal domain, federal opinions are also most likely to be cited within two years of their publication, probably reflecting the time scales at which cases are decided in the courts. The Supreme Court opinions, however, do not show an optimal time scale for discovery.

Another interesting difference between domains is in the rate of forgetting. A newly published scientific paper is more quickly forgotten than a newly granted patent or Supreme Court opinion. Probability of first citation to a scientific paper decreases at a rate $\approx \Delta t^{-2.3}$ (dashed line in Fig.~\ref{fig:first-cite-dist}(a)), while a patent citation probability decreases at a rate $\approx \Delta t^{-1.4}$ (dashed line in Fig.~\ref{fig:first-cite-dist}(b)). The steep drop off in patent citation after about 300 months is an artifact of the finite data size, which constrains citation period to at most 36 years.
Legal opinions are forgotten less quickly than papers. As we would expect, Supreme Court opinions are forgotten even more slowly (as $\approx \Delta t^{-1.4}$) than other federal opinions ($\approx \Delta t^{-1.8}$).

\paragraph{Citation Age}
\begin{figure*}[htb] %
   \centering
   \begin{tabular}{cc}
   \includegraphics[width=0.4\textwidth]{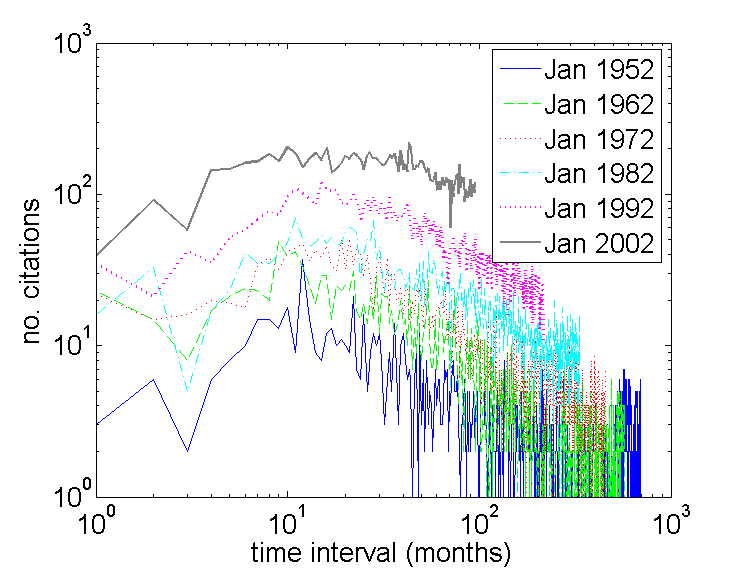} &
   \includegraphics[width=0.4\textwidth]{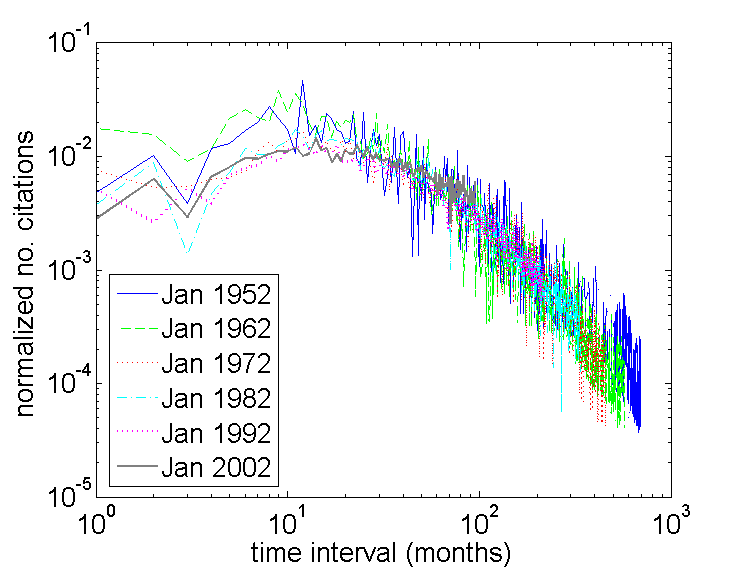}
\\
   (a) & (b)
   \end{tabular}
   \caption{Likelihood of getting cited vs paper's age.  (a) Number of citations made to papers published in the month of January in different decades vs time. An older paper is much less likely to get cited than a younger paper, but more more recent papers appear to get more citations overall. (b) Same data, but  with number of citations made during some month to Jan papers normalized by the  total number of citations made that month.  }
   \label{fig:interval-dist}
\end{figure*}

Figure~\ref{fig:interval-dist} shows how a scientific paper's age affects the likelihood that it will be cited by any paper, not just the first citing paper. Specifically, it shows how the total number of citations received by papers published over a period of one month, here January of different decades, changes over time.
Papers published in more recent decades receive more citations. From this, one may conclude that more recent papers receive more attention, e.g., papers published in 2002 get ten times as many citations after one year as those published in 1952. However, this is simply an artifact of the increasing number of publications. When we normalize the number of citations made during some month to the January papers by the total number of citations made by all papers published that month, the curves line up (Fig.~\ref{fig:interval-dist}(b)). In other words, the fraction of \remove{all citations made to} attention received by a paper $\Delta t$ months after its publication is the same, regardless of when it is published. {A similar normalizing trend was observed by \cite{Radicchi08unversality} when comparing citation rates between different scientific disciplines.}

Attention (via the proxy of citations) decays rapidly in time. This decay does not indicate some sort of rapid obsolescence of scientific research, becoming irrelevant within 12 months of publication. A more likely explanation is that by the time a paper is one year old, it must compete for researchers' limited attention with a rapidly increasing number of newer publications. Later in this paper we examine the mechanisms researchers use to allocate their attention among papers which could account for these trends.

\paragraph{Time since Last Citation}

\begin{figure*}[htb] %
   \centering
   \begin{tabular}{@{}c@{}c@{}}
      \includegraphics[width=0.4\textwidth]{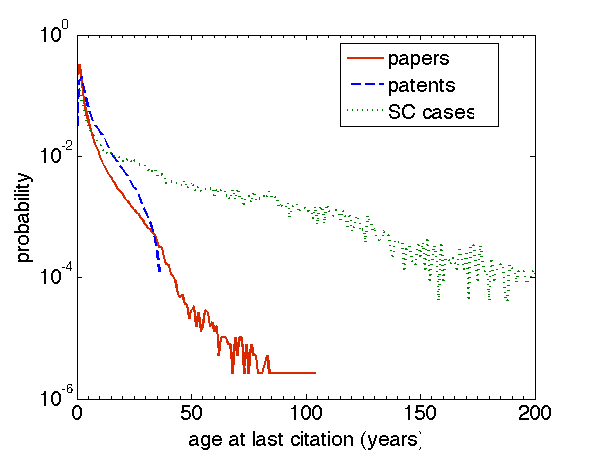} &
   \includegraphics[width=0.4\textwidth]{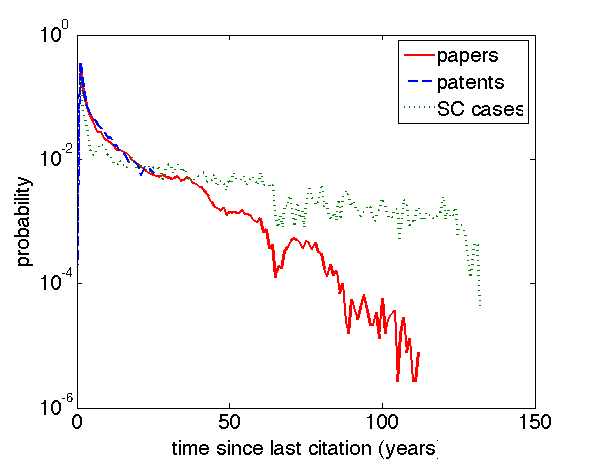}
  \\
   (a)  & (b)
   \end{tabular}
   \caption{Temporal characteristics of the last citation, including the (a) age and (b) time since last citation of a physics paper, a patent and US Supreme Court opinions (SC cases). The time to last citation is calculated from the end of the data set.}
   \label{fig:last-cite-dist}
\end{figure*}

How quickly is information forgotten? Figure~\ref{fig:last-cite-dist}(a) shows the age of the document at the time of last citation. In all three data sets, there exist similar preferences for citing more recent documents, with one- or two-year old documents most likely to be cited. However, the propensity to cite older documents is very different in the three domains. While an older patent is somewhat more likely to be cited than an older physics paper, an older Supreme Court opinion is vastly most likely to be cited.

Differences also emerge when looking at the time since last citation (Fig.~\ref{fig:last-cite-dist}(b)), which give the time scales of obsolescence of information in different domains. Physics papers and patents look remarkably similar to each other, at least over the length of the patents data set. In both APS and Supreme Court data sets, there is a sharp drop in citation probability around 60 years. After that, it becomes increasingly less likely to find a physics paper that has not been cited in decades, while there is a bigger constant probability to find a Supreme Court opinion that has not been cited in decades. These trends could potentially be explained by individual's memory, since sixty years is roughly the span of a scientist's or a judge's career. We leave this question for future research.

\subsection{A Model of Knowledge Discovery in Networks}
\label{sec:models}
In order to cite a work, a scholar must first discover it.
Before the age of the internet, information discovery was a cumbersome process. Scientists discovered new papers outside of the citation network by browsing copies of journals on library shelves, reading popular press, receiving reprints from colleagues, or hearing about them at meetings. Innovators and legal scholars read official government publications to discover new opinions.
The digital age offers many more options for information discovery. Scientists can browse free online resources, such as the \emph{arXiv} preprint server or Google Scholar. Modern search engines index the text of documents, making them discoverable through keyword searches. However, citation networks remain an integral part of the discovery process. As scholars discover works, they may cite them it in their own publications. These new documents may themselves be discovered and read, providing a path for others to discover the original work. 

\begin{figure} [!ht]
   \centering
      \begin{tabular}{c}
   \includegraphics[width=0.5\textwidth]{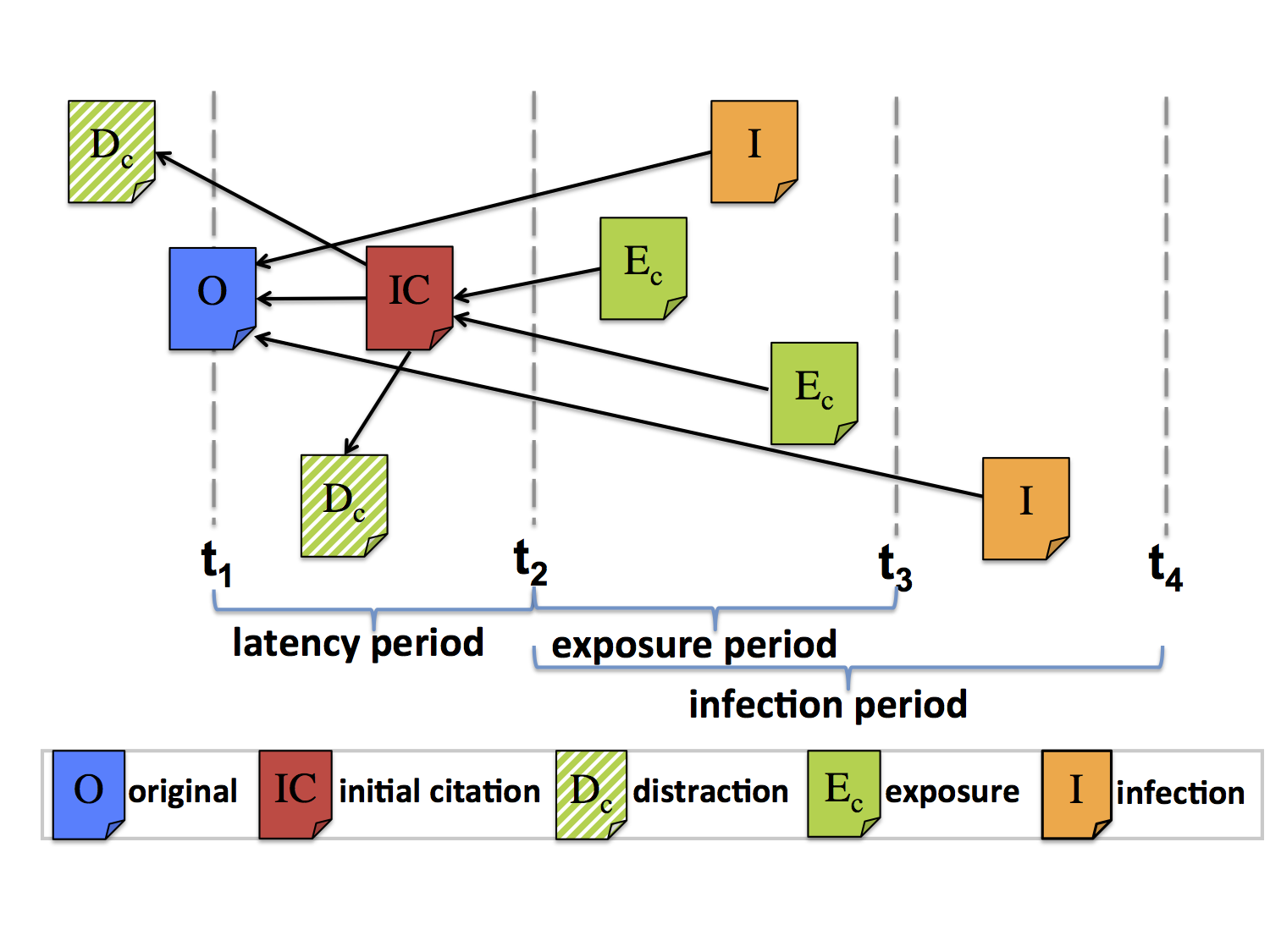} \\
   \end{tabular}
 \caption{A model of knowledge discovery in citation networks, where users discover the original document through citation links from other documents. \label{fig:fsm-schematic}}
\end{figure}

Figure~\ref{fig:fsm-schematic} illustrates the process through which people discover and cite some document published at time $t_1$. We refer to this document, shown in blue in Fig.~\ref{fig:fsm-schematic}, as the \emph{original document}. During some time period $\Delta t_{lp}=t_2 - t_1$, the \emph{latency period}, the original document is discovered and cited by newer documents. We call these citing documents, shown in red, \emph{initial citations}. The red documents are, in turn, discovered during an (arbitrary) \emph{exposure period} $\Delta t_{exp}=t_3 - t_2$. People who find the red documents will be exposed to the original blue document by seeing it cited in the red document. If the original document is relevant, they may then cite it. We call the new citations of the original document \emph{infections}, and they are shown in orange. However, we do not know how these authors discovered the original document. The first infection shown in the figure could have discovered the original document through a red document or another document citing it. In order not to undercount such infections, we make the \emph{infection period} $\Delta t_{inf}=t_4 - t_2$ overlap the exposure period.

Initial citations (red documents) expose readers to the original document. However, each red document may cite many other documents (striped green), which may distract readers from the original document. Hence, we call these other documents \emph{distractions}.

\subsection{Cognitive Heuristics}
The relationships between new citations (infections) and initial citations, exposures, and distractions reveal the processes of knowledge discovery in citation networks and the role of cognitive heuristics in these processes.

\paragraph{Visibility Bias}
We characterize how easily a document can be discovered by its \emph{visibility}. A high visibility document takes little effort to find, because it is seemingly `everywhere.' Visibility may trigger cognitive processes associated with the availability heuristic, which may increase document's perceived importance. Supporting this idea, previous studies have shown that being featured in the popular press~\cite{Phillips91} or the top of \emph{arXiv} daily listings~\cite{Dietrich08,arxiv-visibility}, leads to more citations for scientific papers. In this study we examine the more generic aspect of visibility: being cited by others. Specifically, we ask whether being cited by widely read documents leads to more citations (new infections) for the original work? Every citation chain---solid green $\to$ red $\to$ blue in
Fig.~\ref{fig:fsm-schematic}---raises the visibility of the original document. Without these chains, readers must discover the document \emph{de novo}, so the chains provide a way to quantify boosted visibility relative to other documents of the same age.
We formulate our claim as follows:
\begin{quotation}
\noindent \emph{Visibility bias: documents that receive more exposures \remove{(i.e., easier to find) during the exposure period} will receive more citations during the infection period.}
\end{quotation}

\begin{figure} \centering
\centering     
{   \begin{tabular}{@{}c@{}c@{}c@{}c@{}}
   \includegraphics[width=0.25\textwidth]{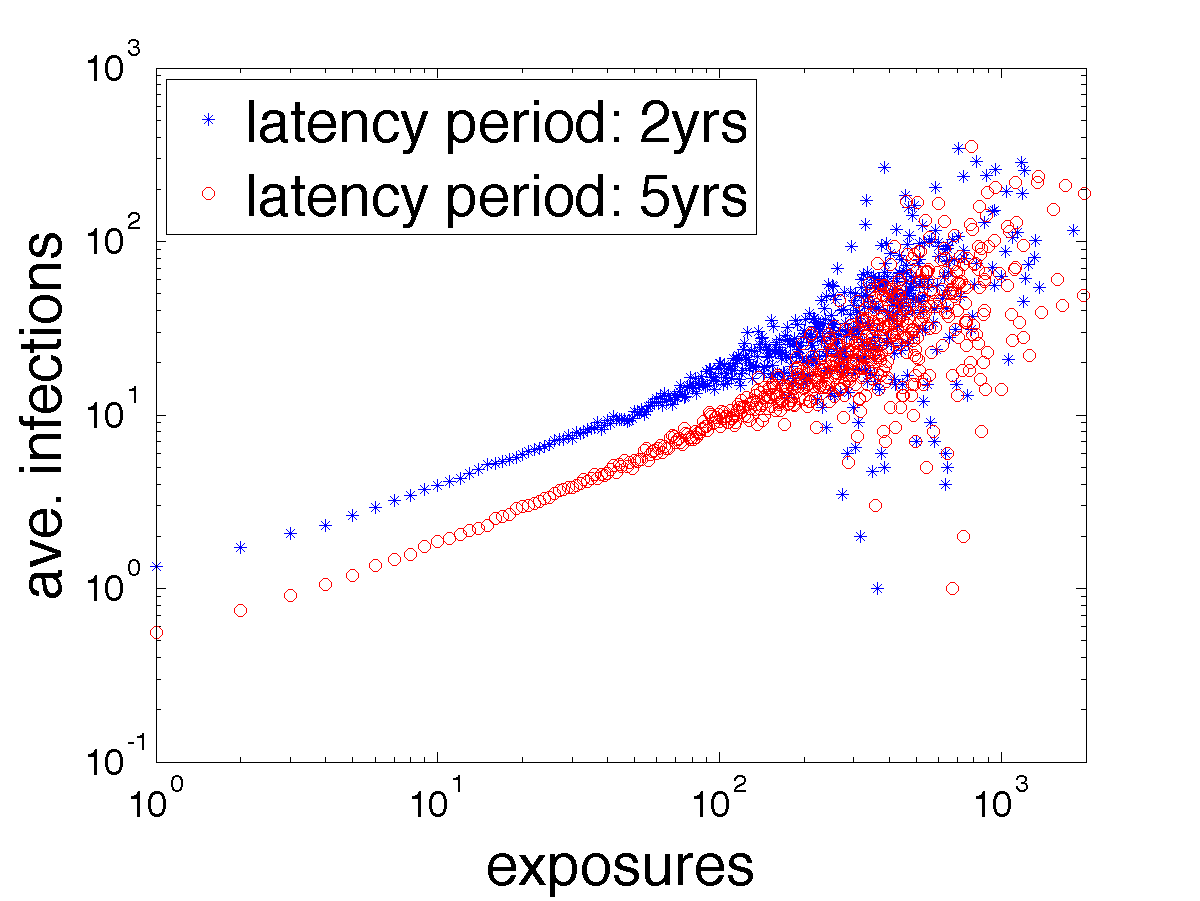} &
   \includegraphics[width=0.25\textwidth]{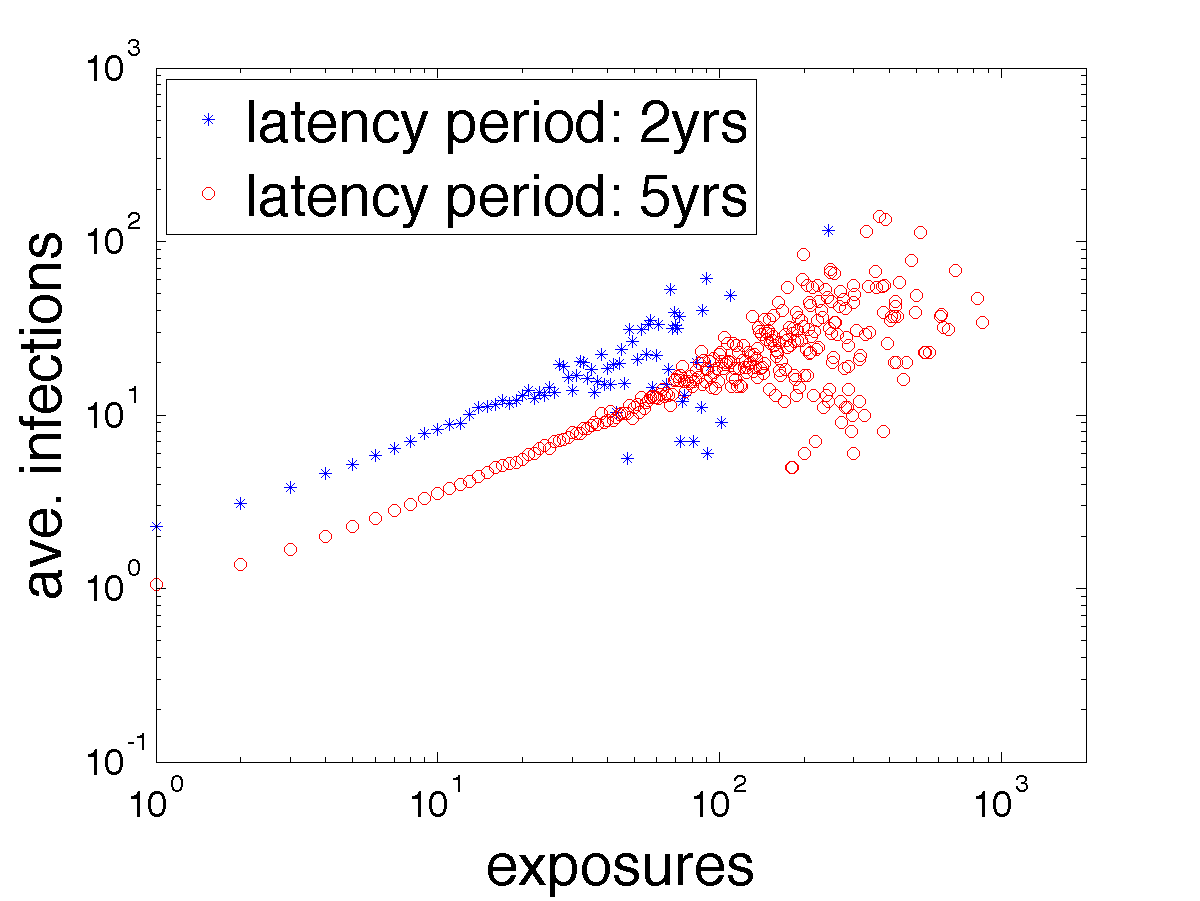} &
   \includegraphics[width=0.25\textwidth]{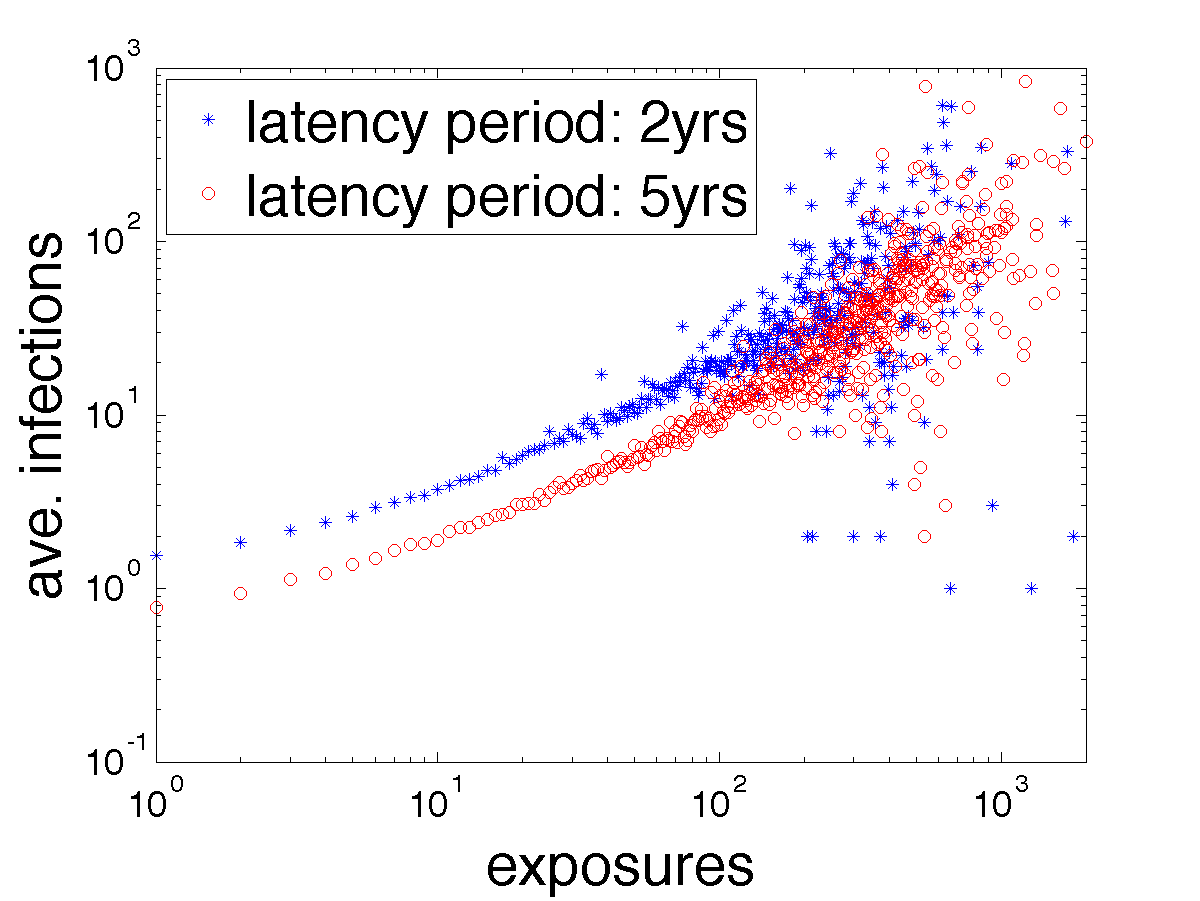} &
   \includegraphics[width=0.25\textwidth]{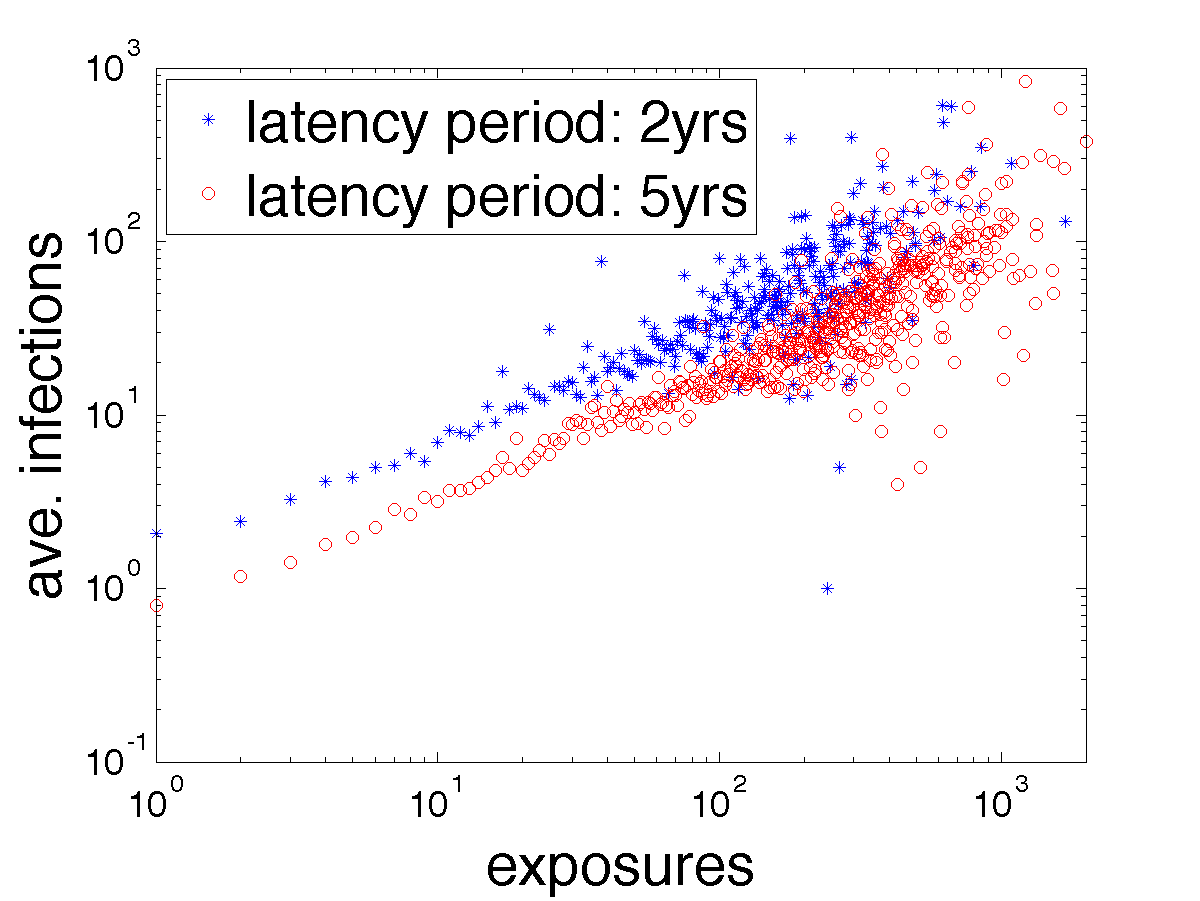} \\
(i) papers & (ii) patents & (iii) opinions &(iv) SC opinions
\end{tabular} }
\caption{\emph{Exposure response}. Average number of new citations (infections) documents receive for a given number of exposures. Exposure period is a two/five year interval following the latency period.}    \label{fig:citations-exposure-curve}
\end{figure}

\begin{table} [!t]
  \centering
  \begin{tabular}{|r||c|c|c||c|c|c|}
    \hline
& \multicolumn{3}{|c||}\emph{Pearson coefficient}	&\multicolumn{3}{|c|}{\emph{linear regression coefficients}}\\ \hline
data (latency period)    &$r(IC,I)$ &$r(E_C,I)$ &$r(a*IC+b*E_C+c,I)$ &$a$ &$b$ &$c$ \\ \hline
papers (LP=2yrs) &0.7147 &0.6852 &0.7410 &0.4271 &0.0665 &0.6026 \\
papers (LP=5yrs) &0.6819 &0.7208 &0.7451 &0.1036 &0.0574 &0.1385 \\ \hline
patents (LP=2yrs) &0.4369 &0.4162 &0.4821 &0.5825 &0.4493 &0.5612 \\
patents (LP=5yrs) &0.5109 &0.5836 &0.6118 &0.1346 &0.1530 &0.3221 \\  \hline
opinions (LP=2yrs)  &0.6872 &0.4894 &0.6873 &0.7687 &-0.0048 &0.1962 \\
opinions (LP=5yrs) &0.8037 &0.7014 &0.8192 &0.2508 &0.0535 &-0.1396 \\\hline
USSC (LP=2yrs) &0.7149 &0.5515 &0.7234 &1.2026 &-0.1231 &1.7000 \\
USSC (LP=5yrs) &0.8566 &0.8137 &0.8801 &0.2358 &0.0773 &-0.0370 \\\hline
  \end{tabular}
  \caption{Correlation of new citations, that are infections ($I$) with the number of exposures received. We report results for three different models of exposure: initial citations ($IC$), citations during the exposure period ($E_C$), or a linear combination of the two, with parameters obtained via linear regression analysis. 
  All correlations are statistically significant ($p<0.001$).}\label{tbl:visibility-correlation}
\end{table}

Figure~\ref{fig:citations-exposure-curve} shows the average number of new citations received by documents with a certain number of exposures. Each line corresponds to a different latency period. Each curve initially grows linearly (in log scale) but later saturates, evidence of the intuition that the more visible a document was during exposure period, the more new citations it received during the infection period.  However, exposures exceeding certain level have a diminishing return in terms of new citations. One possible explanation is that the field become saturated, when all potential citers know of the document, with no possibility of further citations growth.

We quantify the relationships between new citations (infections $I$) and related variables. Table ~\ref{tbl:visibility-correlation} reports Pearson correlation of the new citations with exposures under three different models of exposure: initial citations ($IC$) during the latency period, exposures ($E$), 
and a linear combination of the two, with parameters determined through least square error estimation of linear regression. Correlations are large and significant, with largest correlation obtained through combination of citations during the latency period and exposure period. These positive correlations are consistent with our hypothesis. Note that in a model of how scientific papers accumulate new citations, Wang et al.~\cite{Barabasi13} took a paper's visibility to be the number of citations it receives. This is a good first-order approximations, though indirect exposures through initial citations of the original paper 
also contribute to paper's visibility.
In conclusion, results of aggregate analysis suggest that, all things being equal, a document that has a higher visibility, because it receives a greater exposure through citations, will be cited more in the future than a lower visibility document.
However, not every exposure is equally effective in attracting attention. Below, we attempt to statistically tease out the indicators of how people allocate their attention across different documents and account for how citation information is utilized.

\paragraph{Novelty Bias}
Recency bias is well documented in psychology, which has shown that people often pay more attention to, and more easily recollect, recent stimuli or observations. In this section we show that people intrinsically favor citing more recent documents, after accounting for visibility.
\begin{quotation}
\noindent \emph{Novelty bias: given equal visibility, people are more likely to cite more recent documents.}
\end{quotation}

\begin{figure}%
   \centering
   \includegraphics[width=0.45\textwidth]{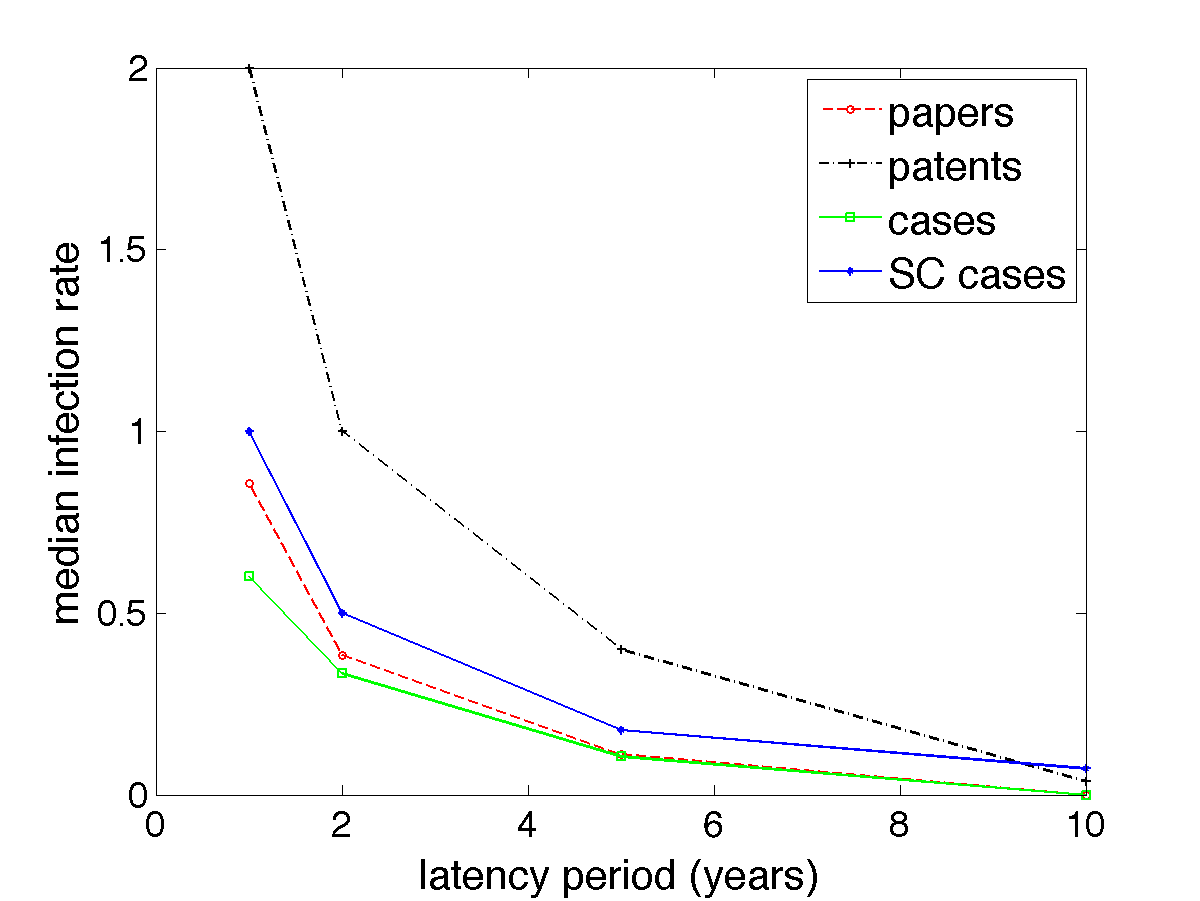}
   \caption{\emph{Novelty bias}. Median infection rate for all documents decreases for longer latency periods, showing preference to cite more recent works.}
   \label{fig:novelty}
\end{figure}

In our data, exposures following a longer latency period are less effective in generating new citations to documents than exposures following a short latency period. To rule out systematic shifts in the populations of documents that receive a certain number of exposures, we study the infection rate, i.e., the rate at which exposures are converted into new citations, on a per-document basis. For each document, we measure the number of exposures it receives during the one-year exposure period following a certain latency period\footnote{We add 1 to all exposures to avoid division by 0.} and the number of new citations (infections) it receives during the infection period. The infection rate for each document for any given latency period is the number of new infections divided by the number of exposures. Figure~\ref{fig:novelty} shows the median infection rate for all documents \emph{vs} the latency period.
The median infection rate decreases over time, evidence of a general novelty bias. 
This finding is consistent with the view that the perceived relevance of scholarly works~\cite{tang2008citation,Barabasi13}--- even of facts themselves~\cite{Arbesman12}---evolves over time as it is refined by successive generations of scientists.

\paragraph{Popularity Bias}
In the absence of information about the costs and benefits of various decisions, people tend to imitate the behavior of others. This cognitive bias, referred to the ``bandwagon effect'' or ``social proof,'' plays a fundamental role in explaining human behavior in psychology, sociology and economics, and is one of the mechanisms for social influence~\cite{Aronson11}. In citations, this bias could manifest itself in a number of ways. Scholars may be more likely to read documents that were cited more frequently. A ranking in Google scholar search results or an endorsement by a digest are some of the signals of a document's popularity.

As a result of the popularity bias, documents that are already popular will receive more attention.  An alternate explanation is that documents that receive more citations are of higher quality, and preferentially receive more citations even in the absence of social signals about their current popularity. We do not attempt to resolve the quality/influence ambiguity in this work, but empirically investigate its presence in our data, referring to it as the ``popularity bias.'' Another possible explanation is that some fields are larger than others, giving documents more potential citations.

\begin{figure}
\centering     
\begin{tabular}{@{}c@{}c@{}c@{}c@{}}
   \includegraphics[width=0.25\textwidth]{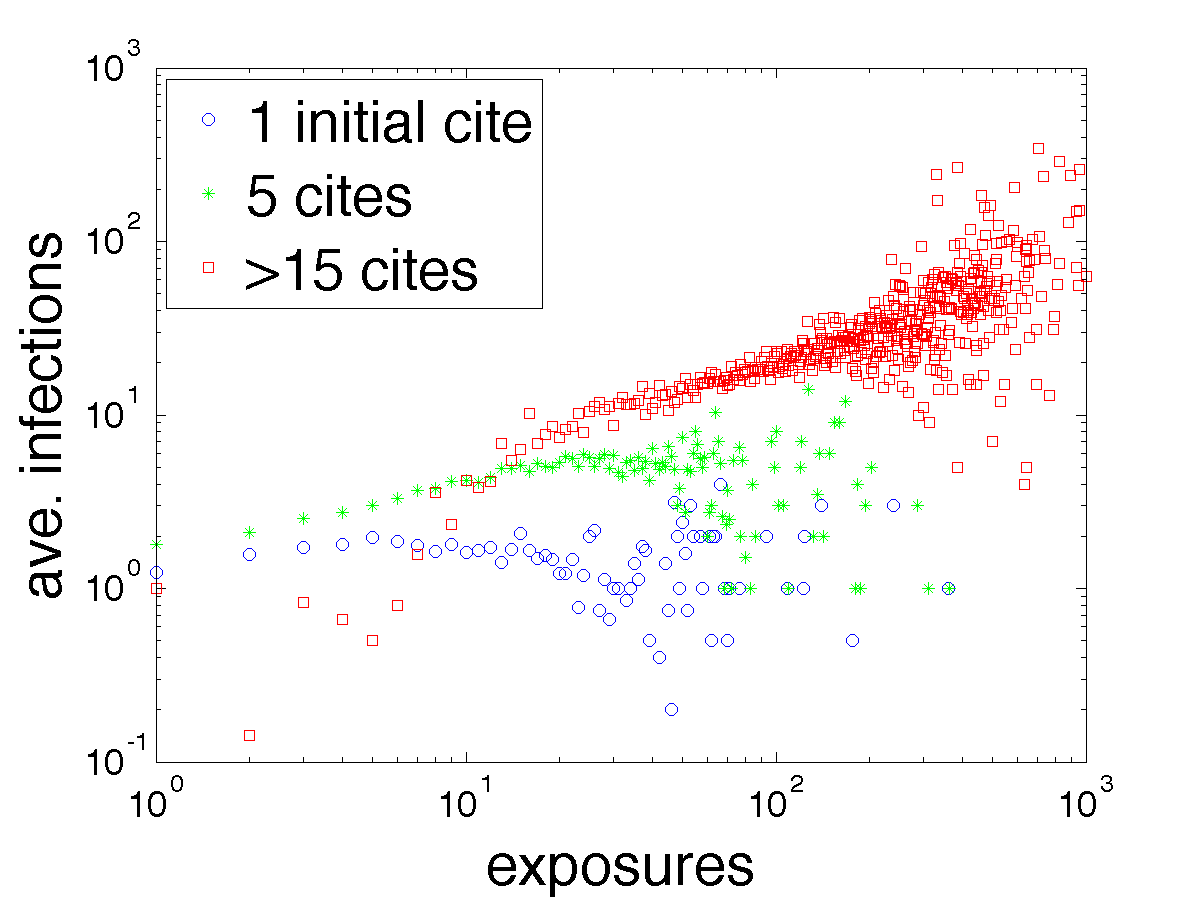} &
   \includegraphics[width=0.25\textwidth]{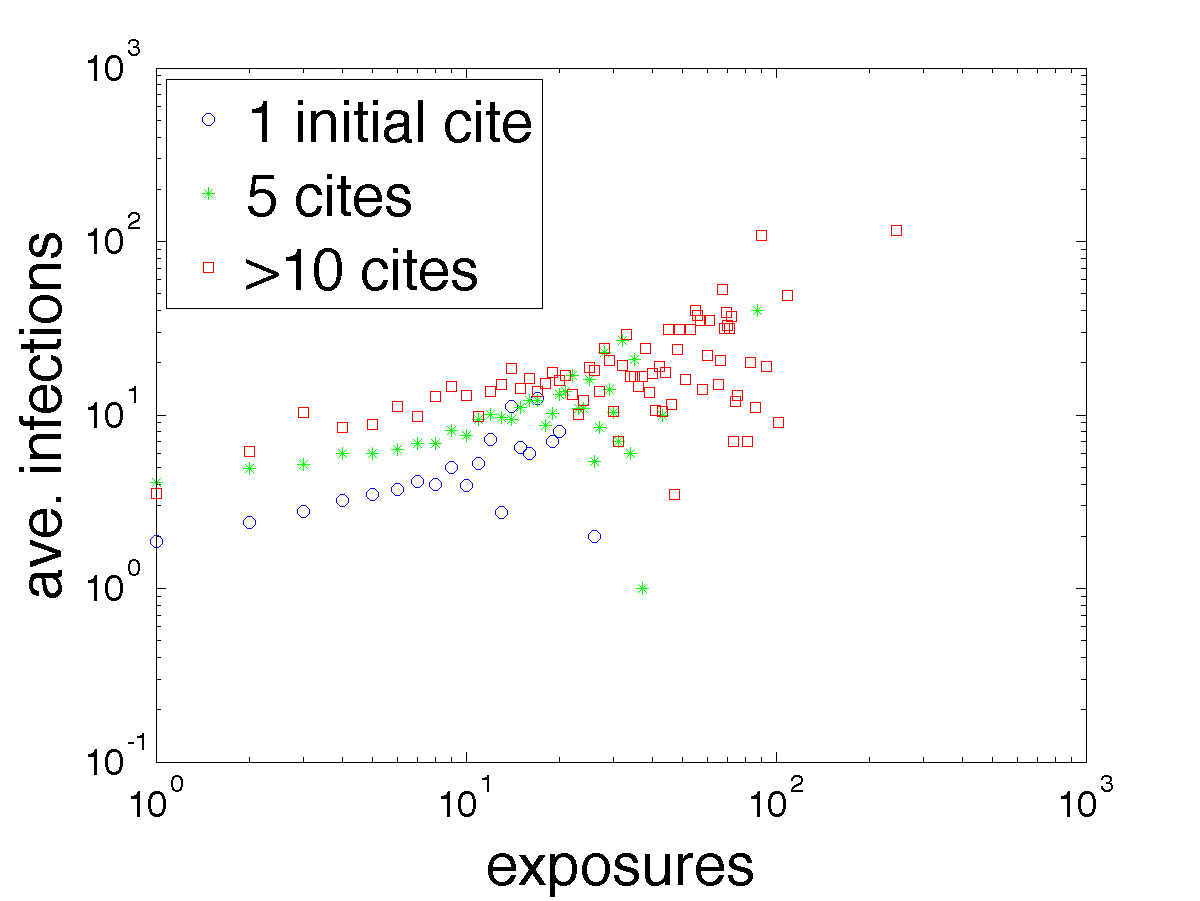} &
   \includegraphics[width=0.25\textwidth]{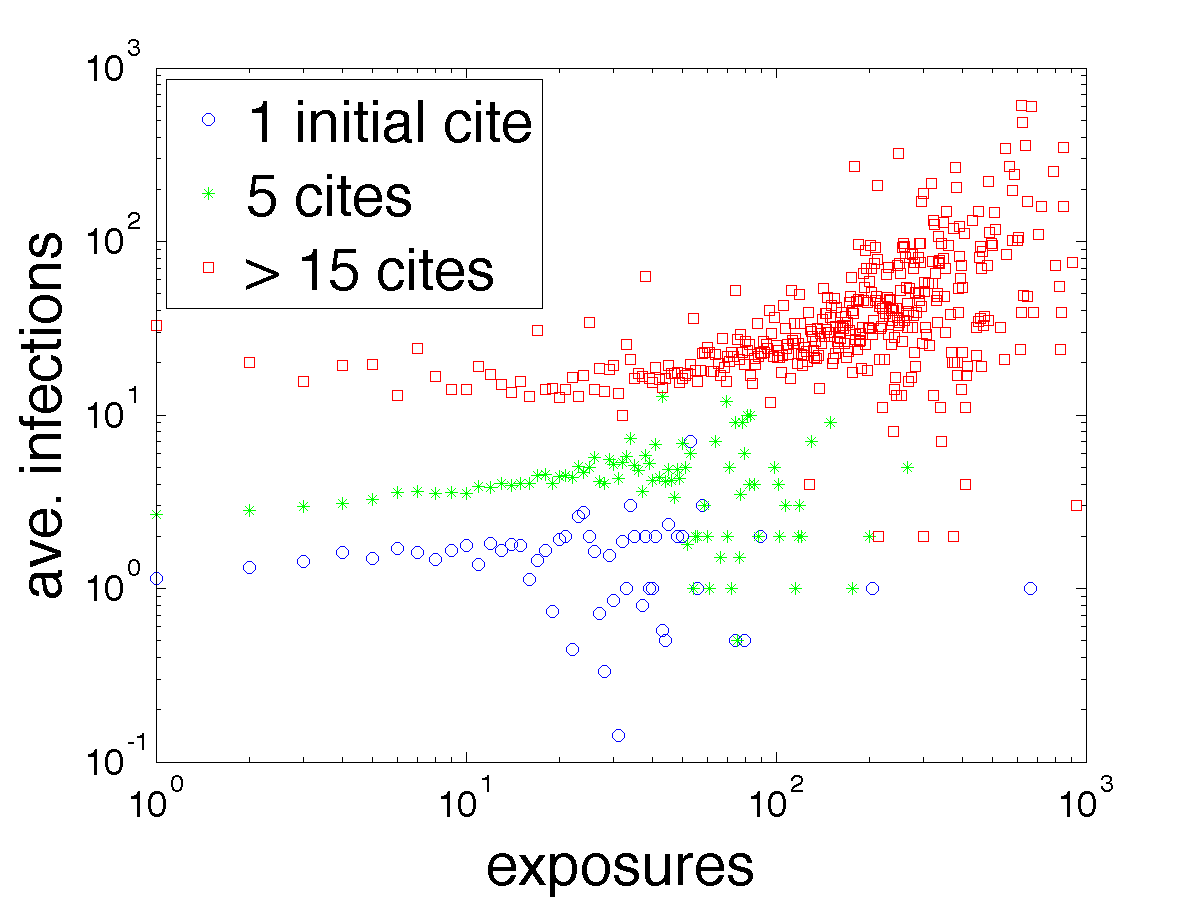} &
   \includegraphics[width=0.25\textwidth]{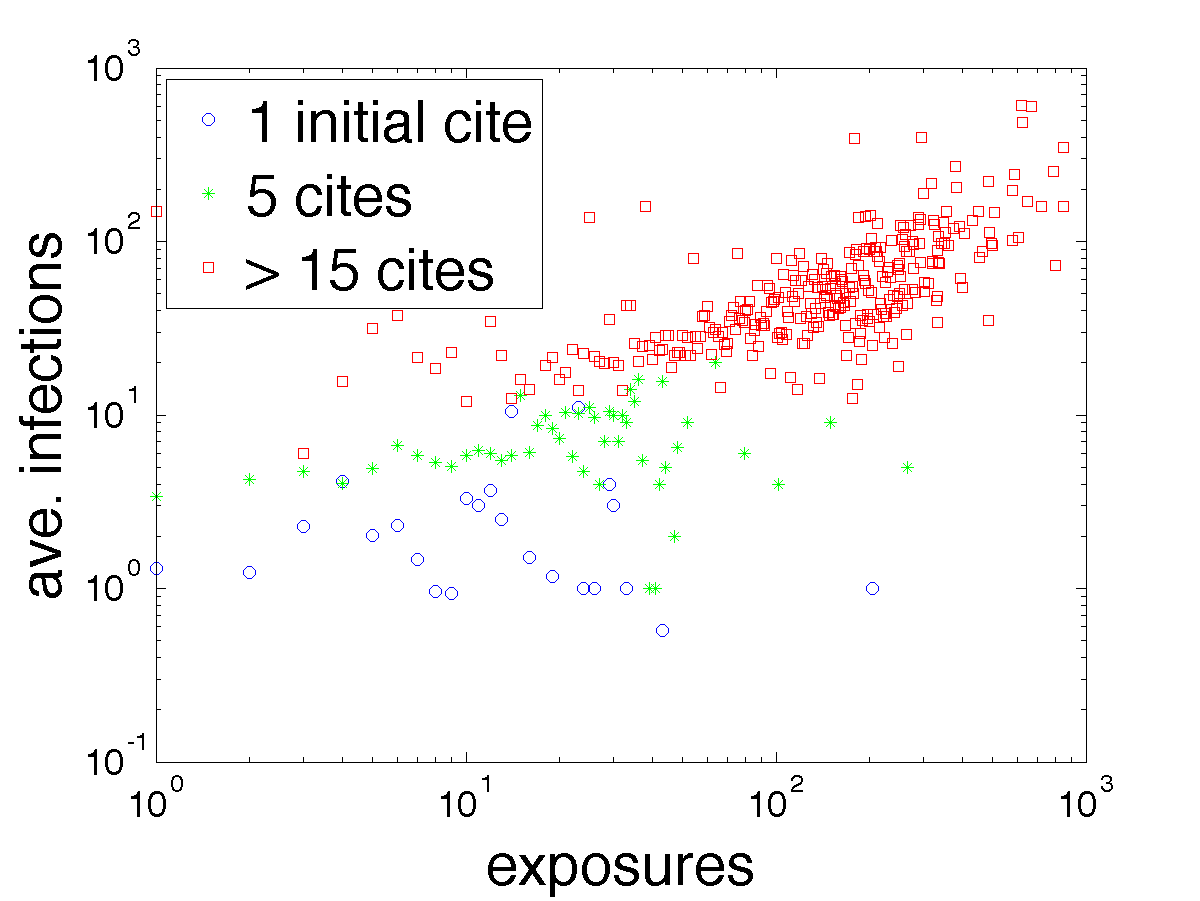} \\
   (i) papers (LP=2yrs) & (ii) patents (LP=2yrs) & (iii) cases (LP=2yrs) &(iv) SC cases (LP=2yrs) \\
\end{tabular}
  \begin{tabular}{|r||c|c|c|}
  \hline
\emph{data (latency period)}  &$MI(E_C, I)$ &$CMI(E_C, I | IC)$ &$II(EC, I, IC)$ \\\hline
papers (LP=2yrs)&0.4362&0.1991&-0.2371 \\
papers (LP=5yrs)&0.4522&0.2772&-0.1751 \\\hline
patents (LP=2yrs)&0.0812&0.0176&-0.0637 \\
patents (LP=5yrs)&0.1994&0.0761&-0.1233 \\\hline
opinions (LP=2yrs)&0.3203&0.0770&-0.2433 \\
opinions (LP=5yrs)&0.3287&0.1036&-0.2251 \\\hline
USSC opinions (LP=2yrs)&0.7600&0.5139&-0.2461 \\
USSC opinions (LP=5yrs)&1.0105&0.6658&-0.3446 \\\hline
  \end{tabular}
\caption{\emph{Popularity bias}. Average number of new citations (infections) received as a function of exposures when documents are separated into different classes based on the number of citations they receive during their first two years. The table shows  Mutual Information ($MI$) between exposures and infections, Conditional Mutual Information ($CMI$) given initial citations and Interaction Information ($II$) between the three variables which is the difference between $MI$ and $CMI$. }    \label{fig:popularity-effect}
\end{figure}

Figure~\ref{fig:popularity-effect} shows the average number of new infections documents receive for a given number of exposures in the three citations domains when these documents are separated into different classes based on the number of citations they received during a two year latency period. These initial citations provide a signal about popularity (or quality) which others may later attend. The classes were defined so as to produce equal statistics bins, whenever possible. They correspond to ``low'' quality documents (one or two citations during the two-year latency period), ``medium'' (3--10 initial citations) and ``high'' (more than 10 initial citations) quality documents.

There exists some separation between the classes for all domains. This means that a medium quality (popularity) document is somewhat more likely to be cited at some exposure level than an low quality (unpopular) document. A higher quality (more popular) document is still more likely to be cited at the same exposure level. Moreover, new citations (infections) saturate sooner for the low quality documents, and never reach the citing levels of more popular documents. Such saturation could be related to the size of the community interested in the document. The larger the community, the greater the citation potential for the document.

The separation between classes is even more pronounced for legal citations than for physics papers. A high quality legal opinion is about ten times more likely to be cited than a low quality opinion at the same exposure. In contrast, there is no separation between the classes of patents. The high quality class has fewer patents in it than other classes, resulting in more noisy signal.

To investigate whether the initial citations account for the correlations between exposures and new citations (infections), we compute the mutual information between exposures and new citations, conditioned on initial citations, and compare this value to the mutual information between the variables. 
Conditional mutual information (Fig.~\ref{fig:popularity-effect}) is significantly less than mutual information in all data sets, 
which indicates initial citations could probably be the common cause of exposures and new citations (infections). This lends evidence to social influence as the source of the bias.

\paragraph{Halo Effect}
Another type of social influence is possible via the ``halo effect''. In cognitive psychology, the halo effect refers to a cognitive bias in which our impressions of a person color our judgements of that person's unrelated character traits, e.g., when physically attractive people are judged to be more trustworthy. In the scientific citation domain, such a bias could manifest itself when readers attribute more importance, and pay more attention, to papers that are cited by important scholars.
Of course, being cited by a highly cited paper will raise the visibility of the cited work, naturally increasing citations. Therefore, we control for the number of exposures of a document and ask whether being cited by few highly cited documents leads to more citations than being cited by many not-as-popular documents.
In other words, is it better to be cited  by one document that receives 100 citations, or by ten documents that receive ten citations each?

\begin{figure} %
   \centering
  \begin{tabular}{@{}c@{}c@{}c@{}c@{}}
   \includegraphics[width=0.25\textwidth]{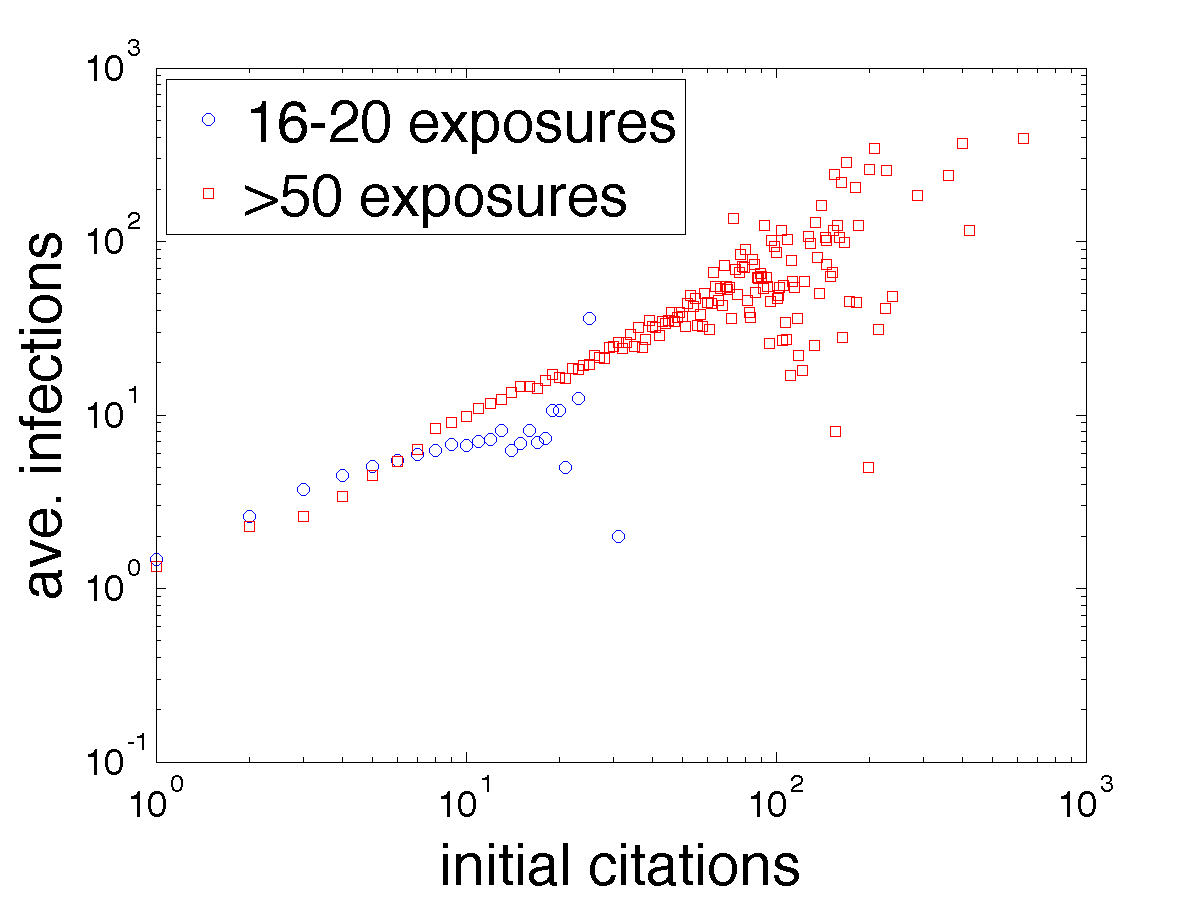} &
   \includegraphics[width=0.25\textwidth]{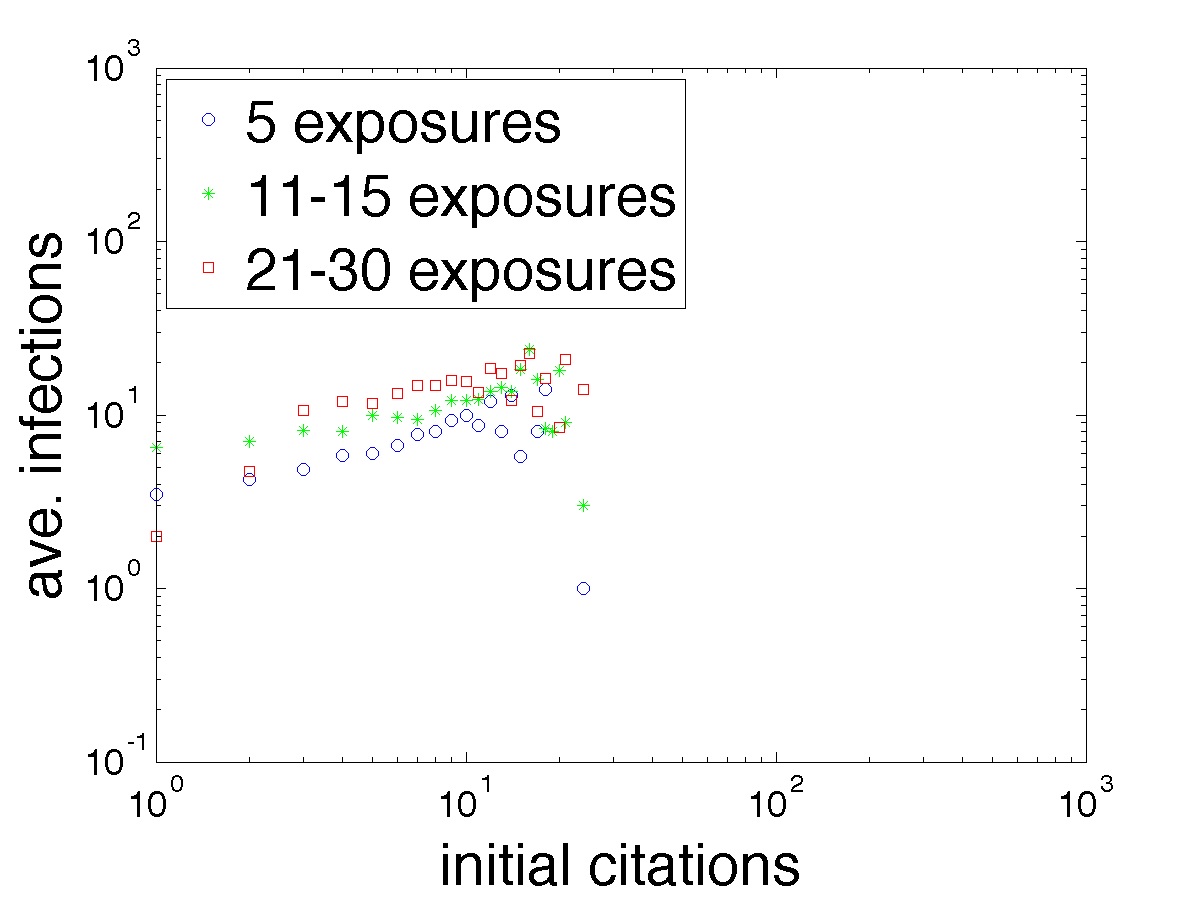} &
   \includegraphics[width=0.25\textwidth]{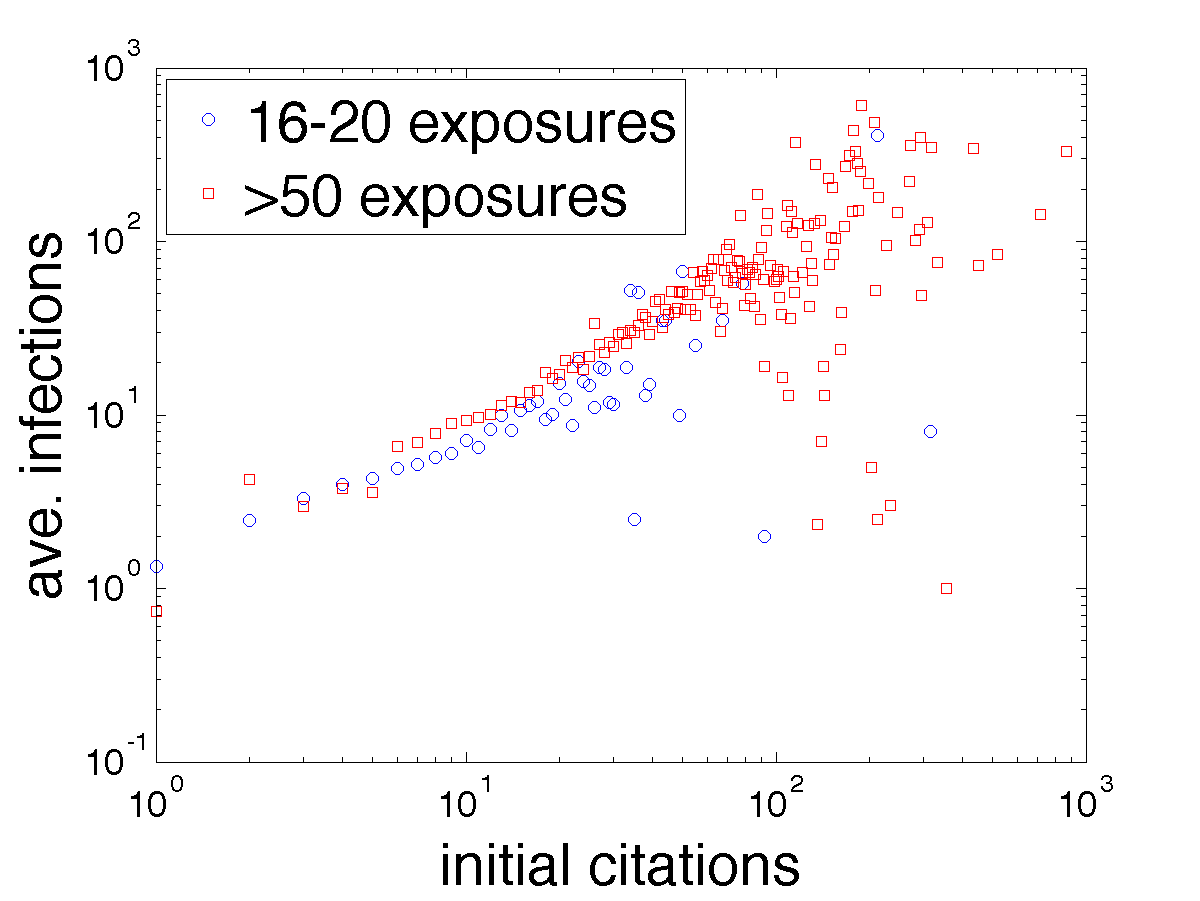} &
   \includegraphics[width=0.25\textwidth]{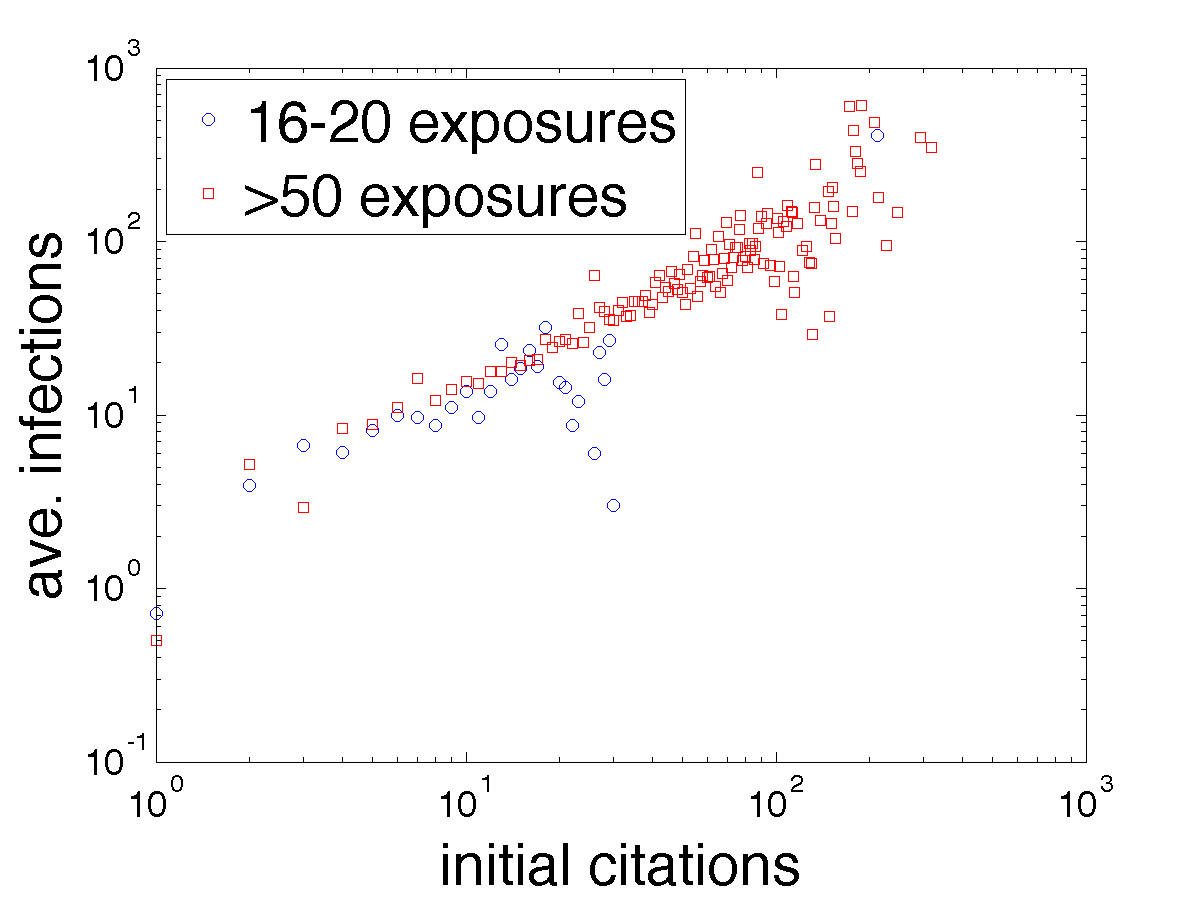} \\
   (i) papers (LP=2yrs) & (ii) patents (LP=2yrs) & (iii) opinions (LP=2yrs) &(iv) SC opinions (LP=2yrs) \\
   \end{tabular}
     \caption{\emph{Halo effect}. Average number of new citations (infections) documents receive as a function of the number of initial citations they received during the latency period when documents are separated into different classes based on the number of exposures they received. For the same level of exposure, the more initial citations during the latency period, the greater the number of new infections. Latency periods are two years ((a)--(c)) and five years ((d)--(e)).}
   \label{fig:halo-effect}
\end{figure}

To investigate the halo effect, we separate documents into classes based on the number of exposures they receive during the exposure period. Figure~\ref{fig:halo-effect} shows the average number of new infections as a function of the number of initial citations these documents received during the latency period. For the same number of initial citations, exposures result in a greater number of new citations (infections), as is expected by papers becoming more visible. If the halo effect held in our data sets, we would see the number of new citations decrease with the number of initial citations for the same number of exposures. However, we see the opposite: for the same number of exposures, having more initial citations leads to more new citations (physics papers and legal decisions) or a constant number of new citations (patents). Consider physics papers that receive exactly ten exposures after a two year latency period (Figure~\ref{fig:halo-effect}(a)). When these exposures come from 10 different papers that cite the original paper during the latency period, that paper will receive seven new citations on average during the infection period. However, if these exposures come from a single paper, the original paper will only receive fewer than two citations on average. The same is true for different number of exposures.

The absence  of an observed halo effect  supports the interpretation of exposures as a measure of visibility, as opposed to a measure of endorsement by popular documents.

\subsubsection{Divided Attention  \label{subsubsection:divided_attention}}
\begin{figure}[htbp] %
   \centering
  \begin{tabular}{@{}c@{}c@{}c@{}c@{}}
   \includegraphics[width=0.25\textwidth]{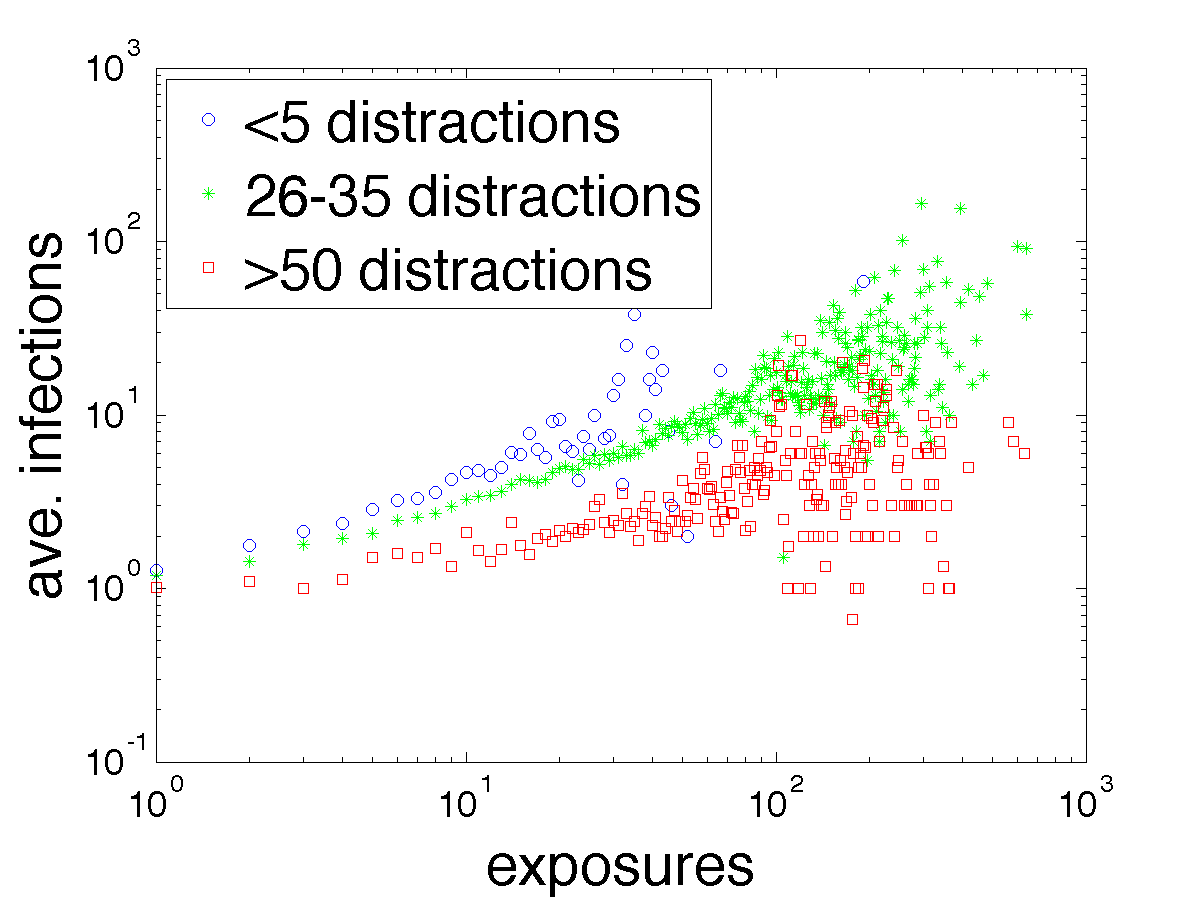} &
   \includegraphics[width=0.25\textwidth]{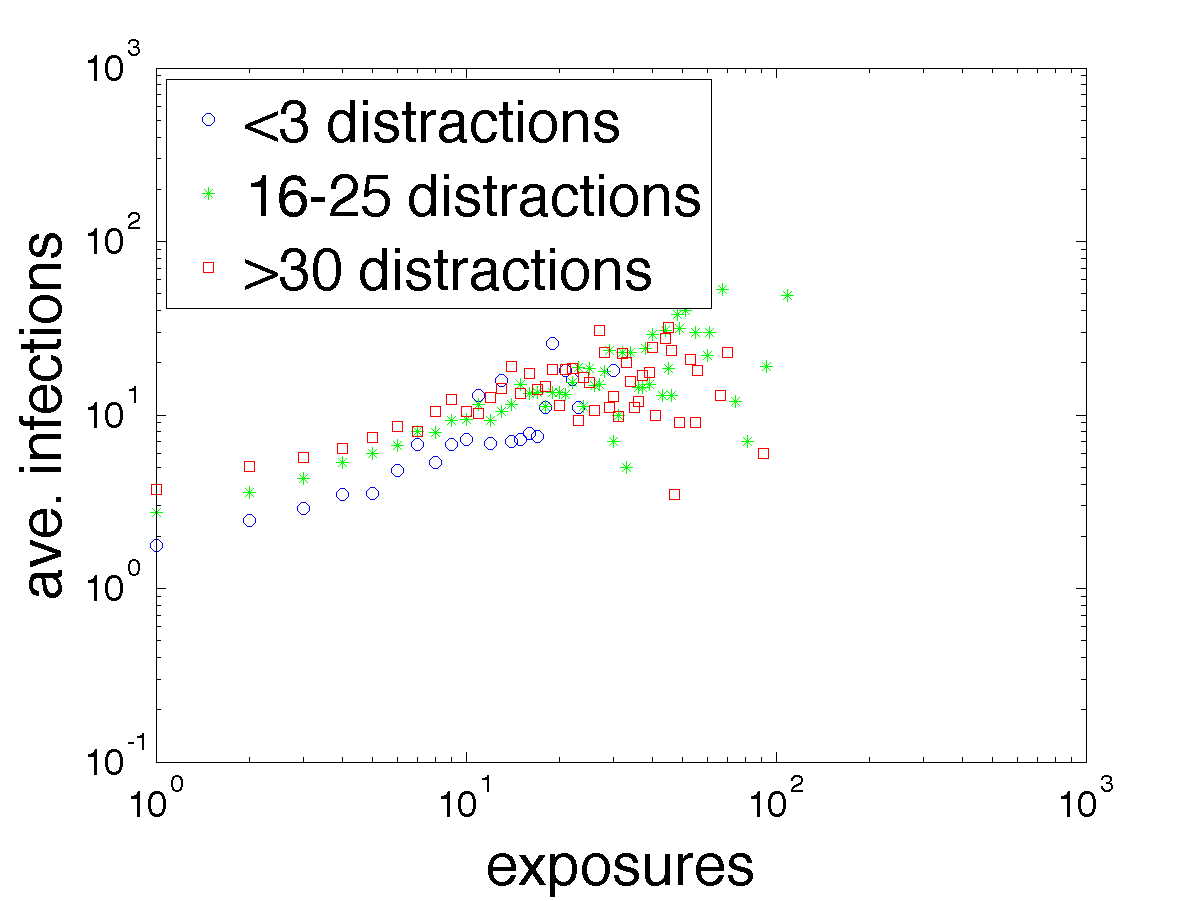} &
   \includegraphics[width=0.25\textwidth]{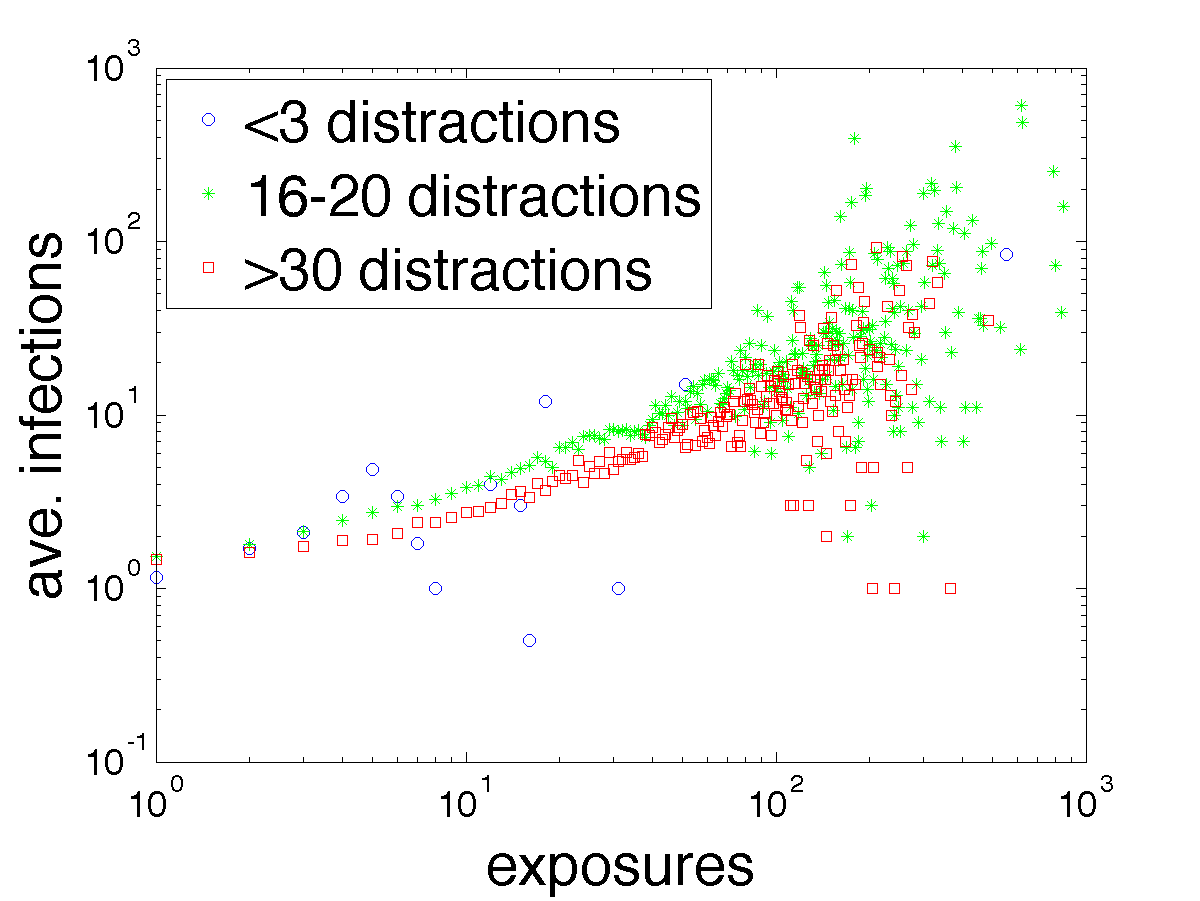} &
   \includegraphics[width=0.25\textwidth]{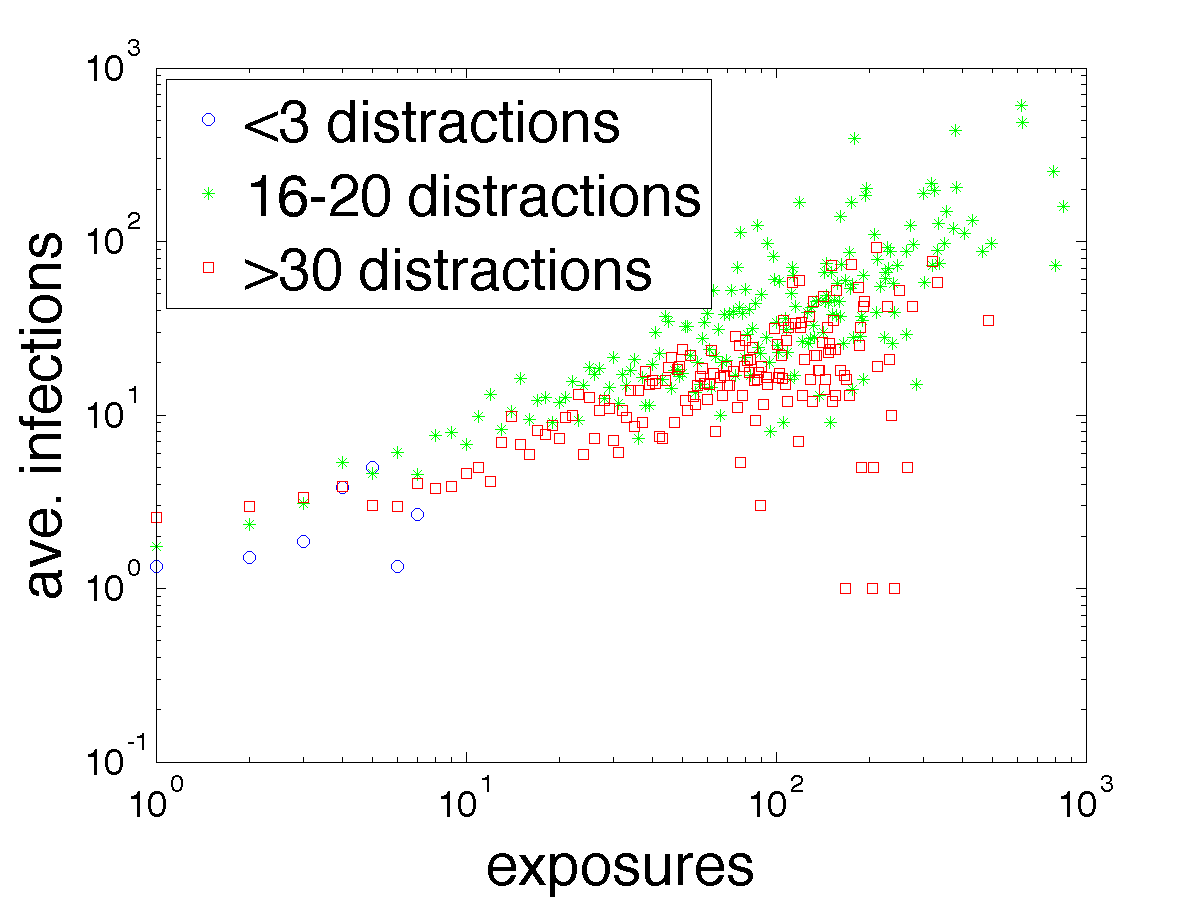} \\
   (i) papers (LP=2yrs) & (ii) patents (LP=2yrs) & (iii) opinions (LP=2yrs) &(iv) SC opinions (LP=2yrs) \\
   \end{tabular}

   \caption{Divided attention. The mean number of references in (red) documents citing the original (blue) document effects the citation rate of the original document.
   Shown is 2 yr and 5 year latency periods. Papers making fewer references in their bibliographies are more effective at generating new citations for their references.
   }
   \label{fig:divided-attention}
\end{figure}

\begin{table}
\centering
  \begin{tabular}{|r||c|c|c|}
  \hline
\emph{data (latency period) } &$MI(E_C, I)$ &$CMI(E_C, I | D_C)$ &$II(EC, I, DC)$ \\ \hline
papers (LP=2yrs)&0.4362&0.4191&-0.0172 \\ 
papers (LP=5yrs)&0.4522&0.4886&0.0363 \\ \hline 
patents (LP=2yrs)&0.0812&0.0336&-0.0477 \\ 
patents (LP=5yrs)&0.1994&0.1293&-0.0700 \\ \hline 
opinions (LP=2yrs)&0.3203&0.1659&-0.1544 \\ 
opinions (LP=5yrs)&0.3287&0.1946&-0.1341 \\ \hline 
USSC opinions (LP=2yrs)&0.7600&0.9385&0.1785 \\ 
USSC opinions (LP=5yrs)&1.0105&1.2073&0.1968 \\ \hline 
  \end{tabular}
   \caption{\emph{Divided attention}. We measure Mutual Information (MI) between exposure and infections, Conditional Mutual Information given distractions and the difference between MI and CMI, which is the Interaction Information (II) between the three variables.
   \label{table:cmi_divided-attention}}
  \end{table}

Readers' limited divided attention could potentially affect a document's visibility. For example, in FSM, if attention is limited, then a document that appears in a long list of references would be allocated less attention because of the distractions from other documents in the list of references, hence would be cited less than a document that appears in a short list of references, simply because readers will not be able to investigate every relevant reference. To demonstrate the phenomenon of divided attention, we study whether being cited by a document with a long list of references (more distractions) leads to as many new citations as being cited by a document with a short list of references (less distractions). Note that we do not know all the citations that appear in the references section, only citations to other documents; however, averaged over all documents, this should be proportional to the number of total citations.
We frame this hypothesis as follows.

%
\begin{quotation}
\noindent \emph{Divided attention: Being cited by a document with more distractions leads to fewer new citations than being cited by a document  with fewer distractions}
\end{quotation}

\begin{figure}[htbp] %
   \centering
  \begin{tabular}{@{}c@{}c@{}c@{}c@{}}
   \includegraphics[width=0.25\textwidth]{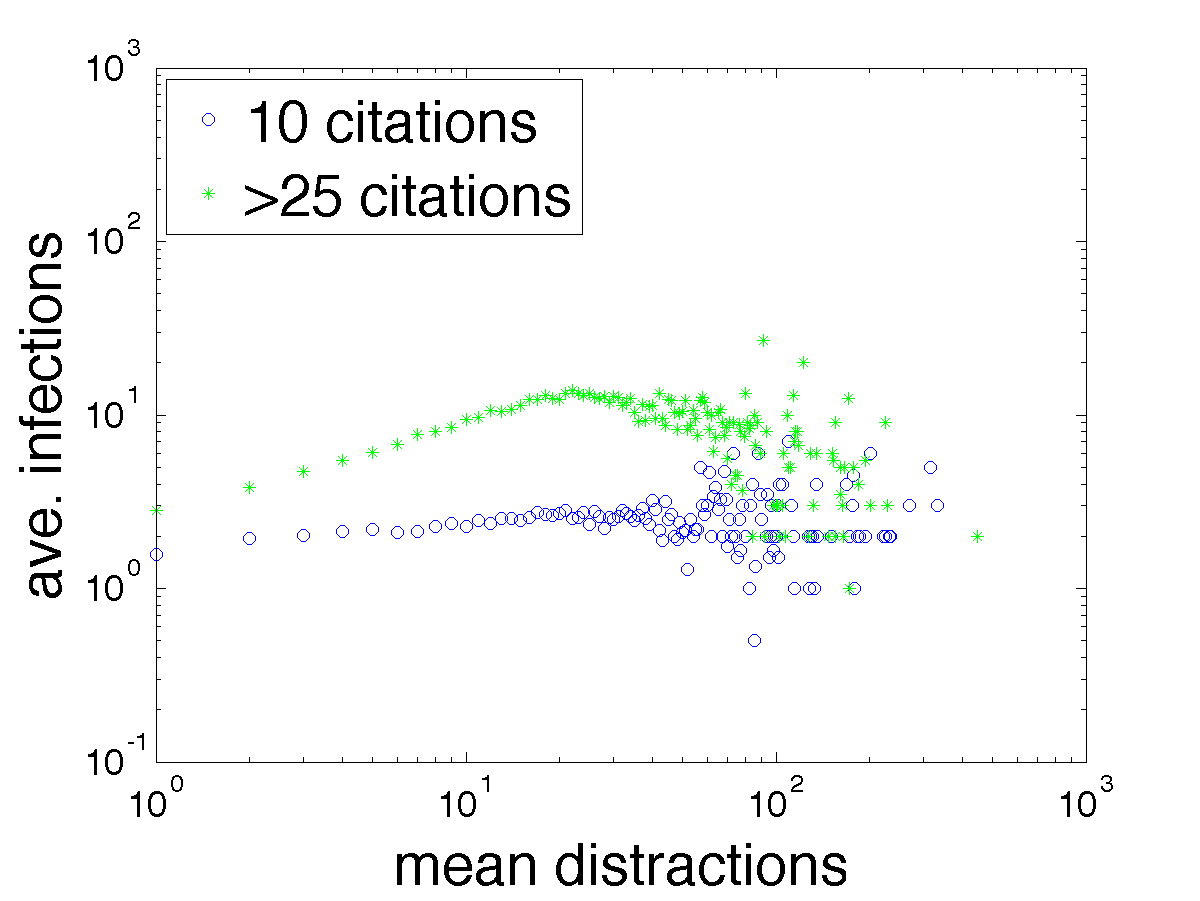} &
   \includegraphics[width=0.25\textwidth]{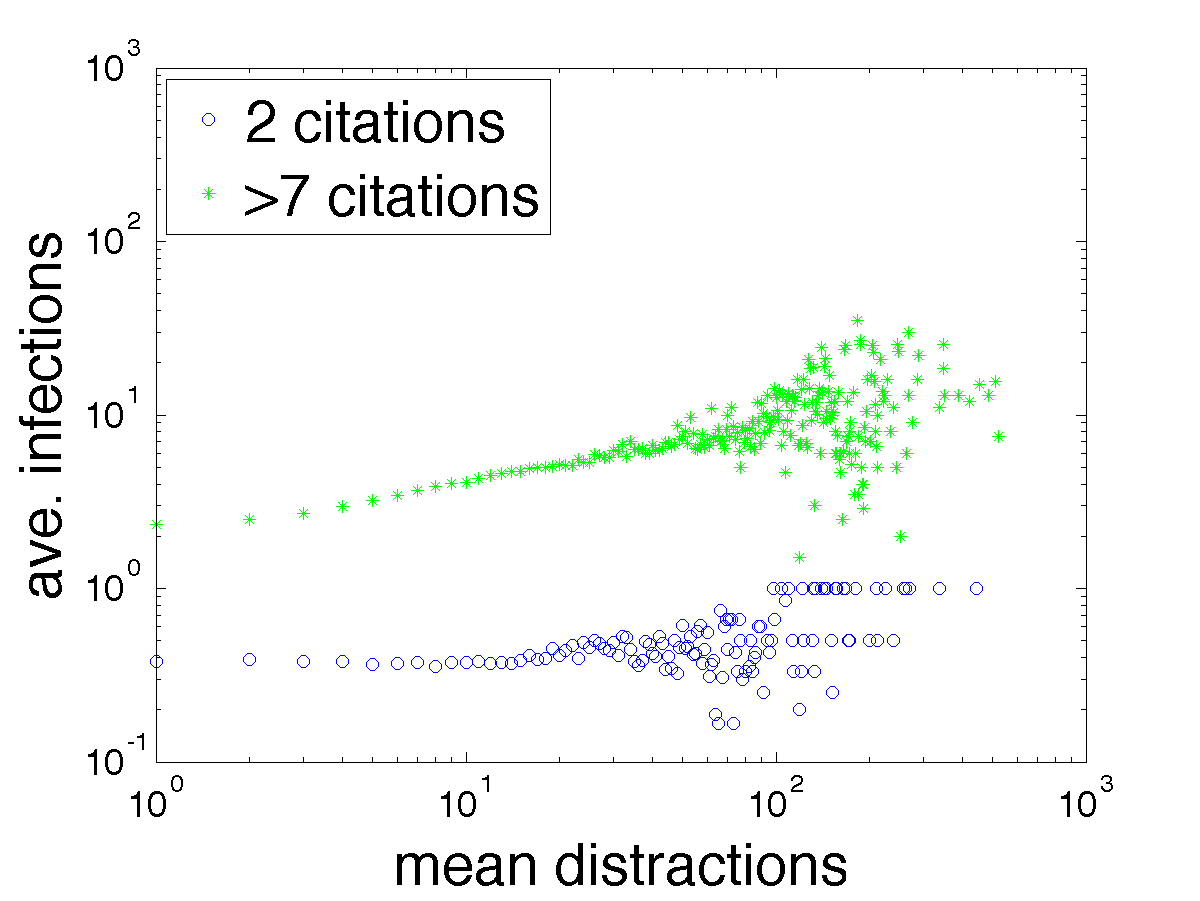} &
   \includegraphics[width=0.25\textwidth]{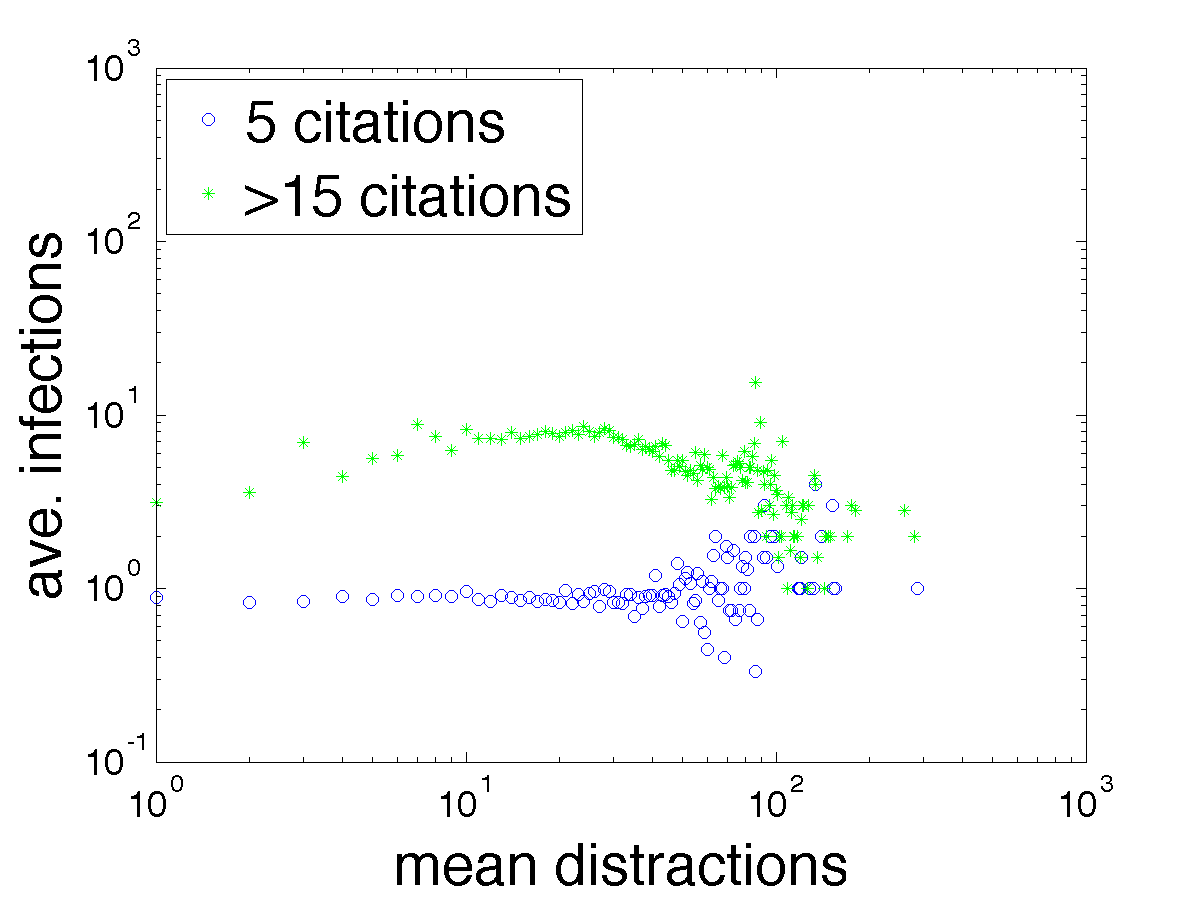} &
   \includegraphics[width=0.25\textwidth]{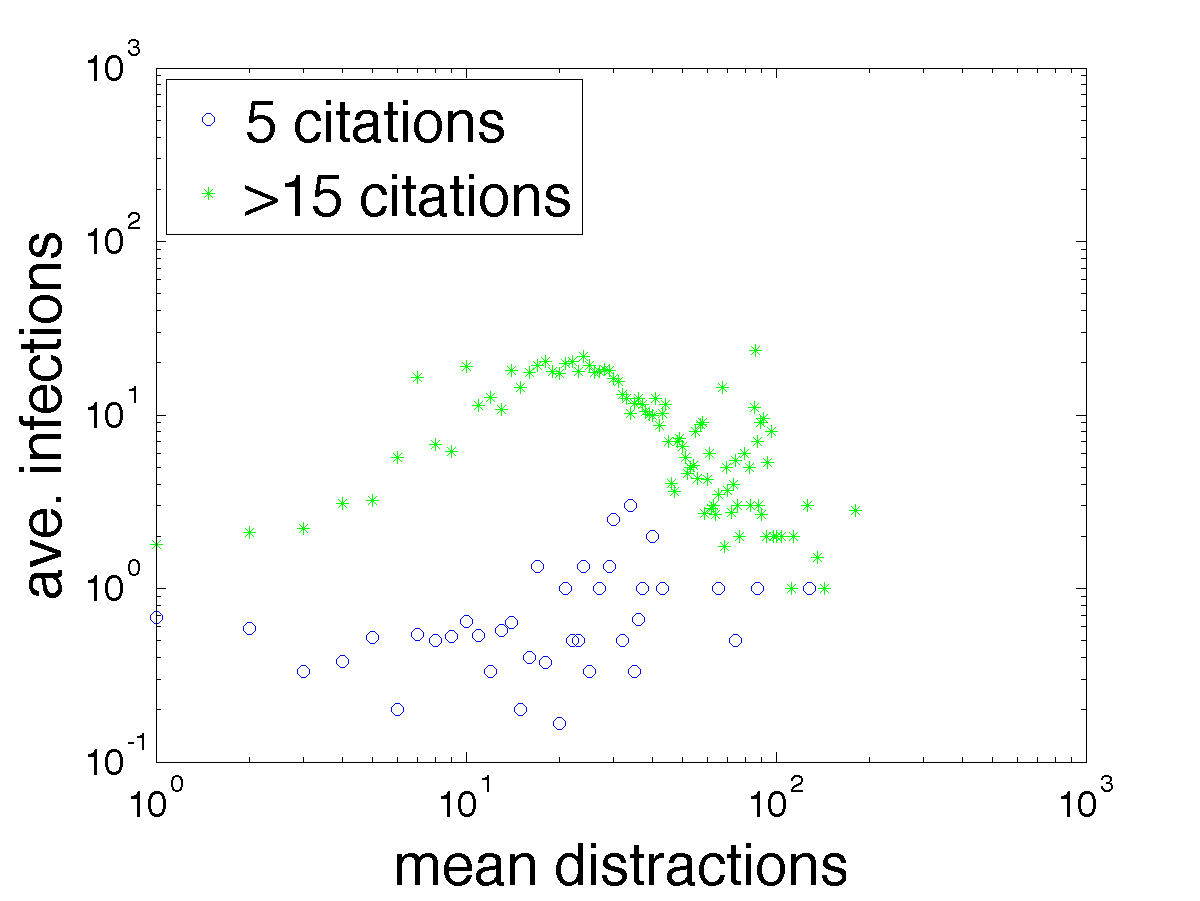} \\
   (i) papers (LP=2yrs) & (ii) patents (LP=2yrs) & (iii) cases (LP=2yrs) &(iv) SC cases (LP=2yrs) \\
   \end{tabular}
   \caption{Breaking down documents into classes with similar quality reveals that divided attention can play a significant role in limiting citation discovery.
   }
   \label{fig:divided-attention_b}
\end{figure}


Figure~\ref{fig:divided-attention} shows the infections vs. exposures plot, with original (blue) documents divided into classes based on the number of mean distractions. The mean number of  distractions is equivalent to the mean number of references contained in (red) documents citing them minus one (rule out the original blue document itself), and the mean number of references is the sum of all references to other documents made by the red documents divided by the number of red documents. We observed that (1) among physics papers and legal opinions, more distractions reduce the likelihood the original document is discovered, as the inflections of a document with low distractions (blue curve) are higher than that with medium distractions (green curve) at the same exposure level, and the infections of a document with medium distractions are higher than that with high distractions (red curve). (2) There is not much evidence of divided attention for the domain of patents and supreme court decisions. This may reflect different conventions within domains: when writing a patent, authors usually exhaustively search for relevant previous patents to prove novelty of the innovation.

Figure~\ref{fig:divided-attention_b} rules out the possibility of biased citation of higher quality (more citable) documents in short bibliographies. We separate the data points into categories with similar quality which are indicated by the total number of citations, and plot average infections vs. mean distractions.
The results show that documents with fewer than ten references in their bibliography section are more effective in generating new citations for their references than documents with more than 30 references, at any level of exposure.
These results are consistent with the phenomenon of divided attention: a cited article ``gets lost'' in a very long list of references, and the longer the list, the easier it is to get lost. Moreover, we observed a turning point in the green curve (low quality documents) that average number of infections increases before this point of mean distractions and drop after it. This suggests a threshold of attention allocation, and when the distractions reach this threshold, attention become divided. However, we did not observe a evident turning point in the blue curve which represents the high quality documents. This may indicate high quality documents would more likely combat or at least alleviate the attention distractions, and not as vulnerable as low quality documents.

\section{Related Work}
Bounded rationality is recognized as an important factor shaping human decisions and behavior~\cite{Kahneman73}, including consumer behavior~\cite{Anderson09,blus:2008tn,Falkinger2007Attention}.
Computer scientists have begun to use attention, a concept related to bounded rationality, to explain social behavior online~\cite{Goldhaber97}, including peer production of content in social media~\cite{Wilkinson08,Hodas14srep}.
Eye tracking studies~\cite{Buscher09,Counts11} and controlled online experiments~\cite{Lerman14plosone} demonstrated the importance of cognitive heuristics and biases in online behavior. Specifically, people pay more attention to content near the top of the page than to lower content, a cognitive bias known as position bias~\cite{Payne51}.
Other studies examined how news articles, Web pages, and YouTube videos, compete for the limited collective attention
~\cite{Moussaid09,Ratkiewicz10,Weng12,Cattuto12}.
Such studies often reproduced the observed macroscopic phenomena, such as the extreme inequality of item popularity, but failed to note the link to individual's cognitive heuristics and biases.

Bounded rationality affects the complexity of social interactions. Dunbar famously found that brain size, and resulting cognitive capacity, imposes an upper bound on the number of social contacts~\cite{Dunbar}.
A study of conversations between Twitter users found that people limit themselves to 150 or so conversation partners~\cite{Goncalves11} and divide their attention among all incoming messages from their friends~\cite{Hodas12socialcom}. Beyond social media, a study of  email communication within an organization  showed that people  located in network positions that expose them to diverse information communicate less with each network neighbor~\cite{Aral07}.
In this paper, we extended these studies and empirically demonstrate that 
cognitive
scientific citation networks.

The citation patterns of scientific articles is of a particular interest to scientists, who have attempted to devise methods to use such data to identify high quality articles,  important emerging trends, or simply track one's own impact.
Merton~\cite{Merton68} described the phenomenon of ``cumulative advantage,'' or the ``rich get richer'' effect, which he argued brought disproportionately greater recognition to scientists who were already distinguished. He explored the possible psychosocial mechanisms of this phenomenon, but as he limited his studies to Nobel laureates, it is not clear whether these principles applied to other scientists.
While Merton did not specifically mention attention, others have  argued~\cite{Klamer02,Franck99} that attention plays a role in the growing inequality of citations scientists receive for their work.  Until now, few studies have explicitly addressed the strategies scientists use to allocate attention. Instead, researchers studied what  features of an article lead to increased citation counts. Such features include the number of authors~\cite{Wuchty07}, coverage by the popular press~\cite{Phillips91}, prestige of the first author~\cite{Merton68}, and journal's impact factor~\cite{Lariviere09}. For example, studies of papers posted to \emph{arXiv} repository  demonstrated a long-term citation boost for papers appearing in the top position of \emph{arXiv} daily announcements~\cite{Dietrich08,arxiv-visibility}.
None of these studies have specifically accounted for divided attention, which prevents unbiased discovery of available literature.
This could create inefficiencies in science, with researchers overlooking some high quality papers or unwittingly duplicating the work of others.


\section{Conclusion}

We described an empirical approach to disentangling factors contributing to the strategies people use to discover knowledge in three different domains: scientific papers, patents, and legal opinions. The key to our approach is to split the data into more homogeneous populations and to carry out `human response dynamics'-based analysis within each population~\cite{Hodas12socialcom}.
Our study identified common patterns in citing behavior. We found that people pay more attention to content that is easier to find: this leads to the \emph{visibility bias}. This bias may result in some documents receiving more citations than less visible documents of higher quality. We also found that attention paid to content decays over time: \emph{novelty bias} leads people to pay more attention to recent information. In addition we demonstrated that \emph{popularity bias} exists: other factors being equal, people prioritize information that received more attention from peers, as measured by the number of citations.
On the other hand, we  found no evidence for another type of social influence: being cited by a popular document does not attract preferentially more attention to the cited document. This is contrary to the halo effect identified by psychologists, in which a person's ability in one area colors the perceptions of others of his abilities in other areas.  While our study does not replace controlled experiments, our analysis of existing behavioral data can guide future experiments to explore the causal relationships proposed here.

These findings suggest that bounded rationality affects knowledge discovery. Rather than evaluate all available documents in order to find the best ones to cite, people rely on simple cognitive heuristics in their citing decisions, for example, whether they saw the document or how new it is.
As the size and complexity of our knowledge continues to grow, the risk introduced by bounded rationality is that high quality relevant knowledge will be missed, thus slowing the pace of innovation. Computational techniques that automatically analyze knowledge contained in citations networks may mitigate this ``burden of knowledge''~\cite{Jones09}. However, the design of such techniques must take bounded rationality into account, or risk exacerbating its effects. For example, ranking content by popularity steers attention to popular content, which may be of lower quality~\cite{Lerman14plosone}. Building systems that help us overcome our bounded rationality should be our priority: our capacity for rapid innovation depends on our ability to quickly discover relevant knowledge in information networks.

\subsection*{Acknowledgments}
This work was supported, in part, by the National Science Foundation (SMA-1360058), and by the Army Research Office (W911NF-15-1-0142), whose support is gratefully acknowledged. Authors are also grateful to the Court Listener project for providing the legal citations data and to Akanxa Padhi for collecting the data used in this project.

\end{bmcformat}

\end{document}